\documentclass[12pt]{article}
\usepackage{titlesec}
\usepackage{authblk}
\usepackage{amssymb}
\usepackage{amsmath}
\usepackage[margin=2cm]{geometry}
\usepackage{graphicx}
\usepackage{fancyhdr}
\usepackage{cite}
\usepackage{subcaption}
\usepackage{enumitem} 
\usepackage{setspace}
\usepackage{xcolor}
\usepackage[dvipsnames, x11names]{xcolor}
\usepackage[format=plain,
            labelfont={bf,it},
            textfont=it]{caption}
\usepackage{xr-hyper}
\usepackage{hyperref}
\usepackage{tikz}
\usetikzlibrary{shapes.geometric}
\usetikzlibrary{shapes}

\newcommand{\mysquare}[1]{\tikz{\node[draw=#1,fill=#1,rectangle,minimum
		width=0.3cm,minimum height=0.3cm,inner sep=0pt] at (0,0) {};}}

\newcommand{\mycircle}[1]{\tikz{\node[draw=#1,fill=#1,circle,minimum
		width=0.3cm,minimum height=0.3cm,inner sep=0pt] at (0,0) {};}}

\newcommand{\mytriangle}[1]{\tikz{\node[draw=#1,fill=#1,isosceles
		triangle,isosceles triangle stretches,shape border rotate=90,minimum
		width=0.3cm,minimum height=0.3cm,inner sep=0pt] at (0,0) {};}}

\newcommand{\mystar}[1]{\tikz{\node[draw=#1,fill=#1,star, star points=6, minimum width=0.3cm, minimum height=0.3cm, inner sep=0pt, outer sep=0pt] at (0,0) {};}}

\hypersetup{
  colorlinks=true,
  linkcolor=magenta,
  citecolor=blue,
  urlcolor=cyan,
  pdftitle={Panda et al.},
  pdfpagemode=FullScreen,
}

\title{\normalsize \vspace{-3em} \textbf{On unconstrained solidification of spherical metallic drops}}
\author[1,3]{\normalsize Priti Ranjan Panda}
\author[1]{Harish Singh Dhami\footnote{Current at Purdue University, West Lafayette, IN, USA}}
\author[1]{Koushik Viswanathan \thanks{Email: koushik\@iisc.ac.in}}
\affil[1]{\textsl{Department of Mechanical Engineering, Indian Institute of Science, Bangalore 560012, India}}
\affil[3]{\textsl{Department of Design \& Manufacturing, Indian Institute of Science, Bangalore 560012, India}}

\date{\normalsize \vspace{-2em} \today}

\newcommand{\rcRO}{\ensuremath{r_c^{RO}}}
\newcommand{\rcCG}{\ensuremath{r_c^{CG}}}
\newcommand{\suppMat}[1]{~\textbf{Supplementary Material,~{#1}~\cite{SuppMat}}}

\titleformat{\section}
{\normalfont\bfseries\scshape}{\thesection}{1em}{}

\titleformat{\subsection}
{\normalfont\bfseries\scshape}{\thesubsection}{1em}{}

\titleformat{\subsubsection}
{\normalfont\slshape}{\thesubsubsection}{1em}{}

\titleformat{\title}{\normalfont\bfseries}{\thesection}{1em}{}

\begin{document}
\maketitle
\vspace{-2em}

\begin{abstract}
  The solidification of metallic droplets into powder particles involves a complex interplay between heat diffusion, surface tension, and geometric constraints. In confined, curved systems—such as those encountered in atomisation, abrasion, and micrometeorite formation—positive curvature and finite boundaries significantly modify classical solidification dynamics. In this study, we systematically investigate the solidification of metallic spheres, focusing on how curvature and confinement influence nucleation pathways, growth kinetics, and interfacial stability. Two competing growth modes—radial outward and circumferential—are analysed using Stefan-type models under a quasi-steady approximation. A generalisation of Mullins–Sekerka stability theory is developed to account for finite spherical domains, revealing that particle size and curvature introduce new destabilising parameters that govern microstructural length scales. Experimental observations of dendritic and cellular morphologies are interpreted through this framework, demonstrating that the interaction between growth fronts, undercooling, and curvature collectively determines the final particle structure. These findings underscore the need to re-evaluate classical solidification theories in the context of curved geometries, with implications for both engineered and naturally occurring metal powders.
\end{abstract}


\clearpage

\section{Introduction}
\label{sec:intro}

\begin{figure}
  \centering
  \includegraphics[width=0.8\linewidth]{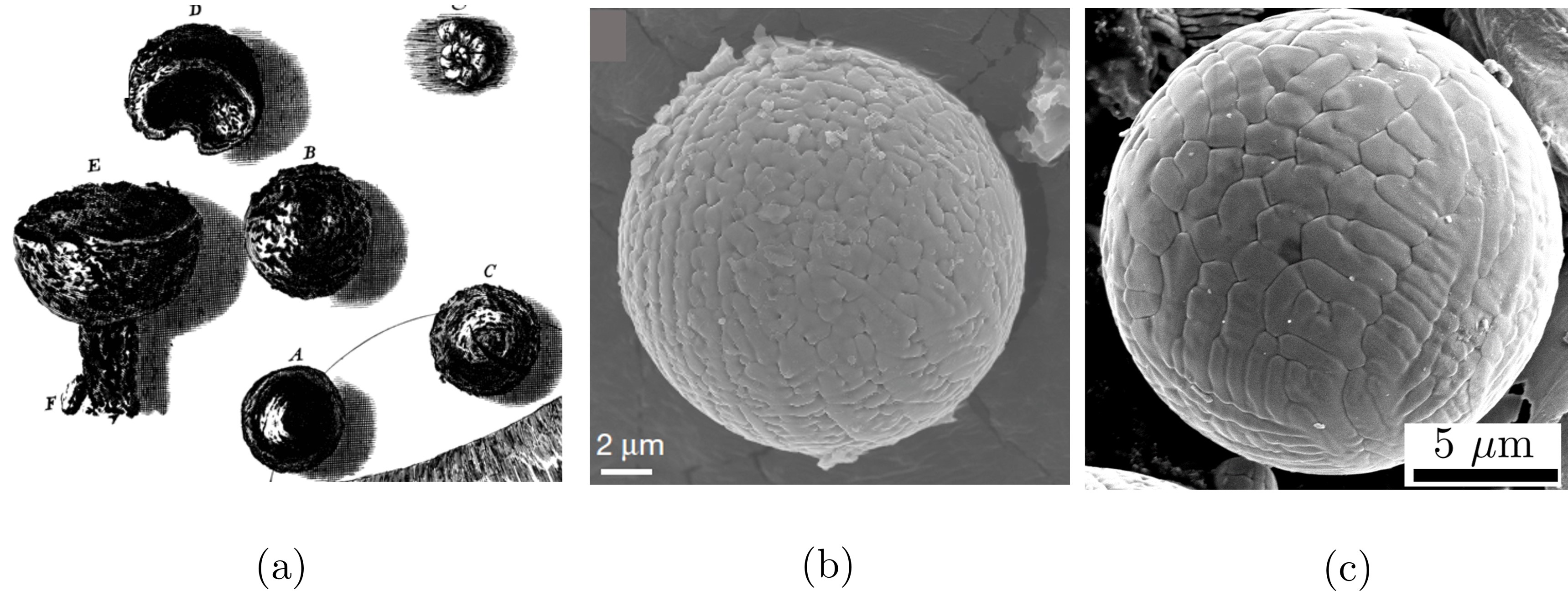}
  \caption{Occurrences of perfectly spherical particles in nature and the characteristic patterns observed on their surfaces. (a) Original illustrations by Robert Hooke showing magnified views of spherical particles produced by striking steel against flint, revealing early observations of solidification morphologies \cite{hooke2007micrographia}. (b) Micrometeorite sample exhibiting a near-perfect spherical geometry with dendritic surface features formed by rapid solidification during atmospheric entry \cite{tomkins2016ancient}. (c) Spherical metallic particles generated via mechanical abrasion, displaying surface patterns indicative of solidification dynamics under curvature and confinement \cite{singh2023fiery}.}
  \label{fig:intro1}
\end{figure}

The large variety of patterns observed during solidification of metals arise from a delicate interplay between heat diffusion and surface tension effects~\cite{winegard1964introduction, mullins1964stability}. The presence of either local curvature or a confining boundary can significantly influence both diffusion and surface tension, resulting in markedly different final microstructures~\cite{langer1980instabilities}. For example, the local equilibrium temperature at the solid--liquid interface is known to depend on curvature, as described by the Gibbs--Thomson relation~\cite{gollub1999pattern}. The curvature of spherical drops modifies the melting point locally and, consequently, the dynamics of the solidification front~\cite{vitelli2006crystallography}. Similarly, confinement by rigid boundaries alters the temperature field, often leading to distinct nucleation and growth behaviours \cite{carslaw1959conduction}. Understanding solidification under these conditions is of considerable practical relevance. Processes such as abrasion and atomisation, commonly used to produce metal powder particles, are ideal examples~\cite{singh2023fiery}. The final particle porosity, surface microstructure and, hence, bulk flowability, are closely tied to the underlying solidification dynamics. Interestingly, similar patterns are observed in natural systems, including micrometeorites (see Fig.~\ref{fig:intro1}), which are shaped by analogous solidification phenomena~\cite{tomkins2016ancient}.

Solidification phenomena in spherical droplets can differ markedly from those in planar, unconfined systems in three very important ways~\cite{horsley2018aspects, vitelli2006crystallography}. Firstly, the high surface-to-volume ratio facilitates extremely rapid cooling rates, allowing for significant supercooling. For instance, homogeneous nucleation in Sn and Bi droplets of approximately $20~\mu$m diameter, produced via emulsification, can persist down to temperatures as low as $\sim 0.3 T_m$, where $T_m$ is the melting temperature~\cite{sheffield1979solidification}. Moreover, the critical size for nucleation on the outer drop surface is highly sensitive to local curvature, and heterogeneous nucleation may occur at significantly higher temperatures~\cite{gomez2015phase, herlach1994non}. Under suitable conditions, growth may initiate earlier on the surface than in the bulk, with attendant characteristic signatures in the final solidified particle.

Secondly, the transition from nucleation to growth of a stable crystal becomes highly size and location dependent. While crystallization phenomena have been explored in curved geometries before, revealing unique non-Euclidean effects such as geometric frustration on the small scale~\cite{ortellado2020phase, meng2014elastic, ortellado2022two}, the  role of simultaneous confinement and curvature on continuum-scale solidification dynamics remains relatively underexplored. Addressing this gap requires systematic comparisons with flat-space analogues to isolate geometric contributions to interface evolution. Confinement and curvature introduce non-trivial modifications to both the solidification time and the resulting surface and internal morphologies~\cite{nichiporenko1968fashioning, levi1982heat}. Experimental studies on solidified droplets, such as those involving Al-Fe alloys, have demonstrated that these effects are strongly curvature dependent~\cite{boettinger1986analysis}.

Finally, an additional complication, particularly prominent in spherical droplets, is the occurrence of multiple simultaneous nucleation events, producing growth fronts that propagate radially outward, inward, and tangentially along the surface. Extensive analytical treatments of radially inward/outward fronts may be found in the literature (see, for instance Refs.~\cite{stewartson1976stefan, gill1981rapid, mccue2008classical} and references therein) using asymptotic expansions and perturbation methods to obtain closed form results for the location of the growing front~\cite{pedroso1973inward,riley1974inward,feuillebois1995freezing}. The  stability of such fronts, however, has received far less attention than their unbounded planar counterparts~\cite{mullins1963morphological, alexandrov2024mullins}.  The possibility of circumferential growth on the surface vis-\'a-vis radially symmetric growth has received much less attention, though it is experimentally prominent. Furthermore, the presence of a finite, curved boundary imposes geometric constraints that influence both the onset and evolution of morphological instabilities, even in the purely radial case, necessitating a re-evaluation of classical stability theories.

In this work, we attempt to address these gaps by presenting a systematic investigation of the solidification of metallic drops. Building on our recent experimental work involving an abrasion-based method, we produce spherical metal particles under controlled conditions, \emph{in lieu} of a conventional atomization based process~\cite{singh2023fiery, dhami2022production}. Despite the inherent spatial complexity, we identify and analyze two distinct, somewhat idealized, solidification modes: one propagating radially outward from the centre of the droplet, and the other advancing along the surface from an initial seed nucleus. We first evaluate the thermodynamics of nucleation in each mode, followed by an analysis of growth front propagation. The stability of these fronts is examined by solving a Stefan-type problem under appropriate boundary conditions. To quantify the role of the finite domain boundary on pattern formation, we extend the classical Mullins--Sekerka stability theory~\cite{mullins1963morphological} to incorporate the finite outer boundary of the particle, thus providing a refined predictive framework for microstructural length scales. Based on our findings, we propose that particle size and the interaction between competing growth fronts---alongside thermodynamic driving forces and geometric constraints---govern the observed morphologies. This work establishes a quantitative foundation for understanding non-equilibrium solidification in confined, curved systems, with broad implications for practically relevant powder production processes.

This manuscript is organized as follows. The experimental details and observed morphologies in solidifying metallic drops are presented in Sec.~\ref{sec:experimental}. We present a comprehensive analysis of the solidification process in Sec.~\ref{sec:analysisResults}, including dimensional considerations (Sec.~\ref{subsec:physicalProcesses}) to outline two distinct, competing solidification modes (Sec.~\ref{subsec:competingsolidificationmodes}) that will form the basis for the rest of the analysis. We then present thermodynamic analyses for nucleation in each mode (Sec.~\ref{subsec:nucleationThermodynamics}), followed by estimates of corresponding growth rates in Sec.~\ref{subsec:growthKinetics}. The stability of growing fronts in each of these modes is evaluated using linear stability analysis in Sec.~\ref{subsec:stabilityGrowthFronts} and the most dominant growth modes are identified. We use the entirety of our analysis to present semi-quantitative interpretations of observed patterns in Sec.~\ref{sec:Interpretation}. A detailed discussion of our analysis and its implications is presented in Sec.~\ref{sec:discussion}, followed by concluding remarks in Sec.~\ref{sec:conclusions}.

\section{Experimental Details}
\label{sec:experimental}

\subsection{Methods}

Spherical metallic particles were generated using an abrasion-based technique, which provides a practical alternative to conventional atomization methods. This approach avoids many of the complexities associated with gas or plasma atomization, yet reliably produces near-perfect spheres, as described in detail in Refs.~\cite{singh2023fiery, dhami2022production}. A brief overview is provided here.

\begin{figure}[h!]
    \centering
    \includegraphics[width=0.8\linewidth]{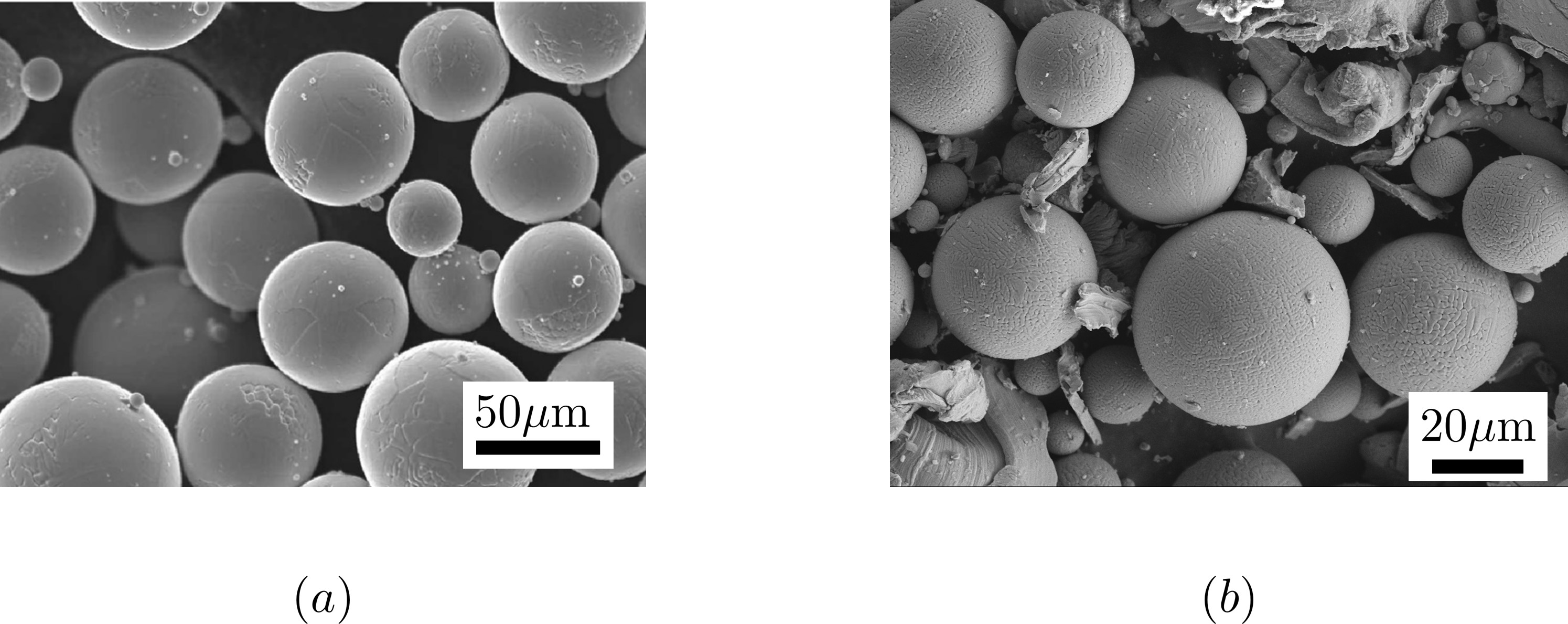}
    \caption{Comparisons of solidified metallic particles produced via (a) plasma atomization~\cite{chen2018comparative} and (b) abrasion. In the latter, stringy chips (fully or partially un-melted) are also visible alongside spherical particles.}
    \label{fig:particleComparisons}
  \end{figure}

An alumina abrasive wheel of diameter $170$\,mm, rotating at $2800$\,rpm, was used to abrade a metallic workpiece (AISI 52100 steel). Material is ejected from the surface and undergoes melting due to oxidation induced heating resulting in the formation of spherical droplets, that subsequently solidify to form powder particles, see Fig.~\ref{fig:intro1}(c). Particle sizes, typically normally distributed between 1 $\mu$m and 150 $\mu$m, can be adjusted by varying the depth of engagement of the wheel with the workpiece. The resulting particles are collected and imaged using a scanning electron microscope ( SEM, Zeiss Ultra55), revealing a range of morphologies. A qualitative comparison between particles produced by abrasion and those formed via plasma atomization is shown in Fig.~\ref{fig:particleComparisons}, highlighting the morphological similarities between the two processes.

The thermophysical properties for the particles (here AISI 52100 steel) are well-approximated by those of pure Fe, and provided in Table~\ref{tbl:materialPropsFe}. After collection, particles were sieved to separate stringy chips and debris originating from abrasive wheel wear. To record the final solidification patterns of the spherical particles, samples were examined under SEM. 
\begin{table}[h!]
  \centering
  \caption{Material parameters for pure Fe~\cite{powell1939further, assael2006reference}.}
  \label{tbl:materialPropsFe}
  \begin{tabular}{l r}
    \hline\hline
    Parameter & Value \\
    \hline
    Melting temperature, $T_m$ & $1813$\,K \\
    Surface energy, $\gamma$ & $2.98$\,J/m$^2$ \\
    Latent heat of fusion, $L_f$ & $247$\,kJ/kg \\
    Density of solid, $\rho_S$ & $7800$\,kg/m$^3$ \\
    Density of liquid, $\rho_L$ & $7000$\,kg/m$^3$ \\
    Thermal conductivity of solid, $k_S$ & $72$\,W/mK \\
    Thermal conductivity of liquid, $k_L$ & $36$\,W/mK \\
    Specific heat at constant volume, $C_v$ & $450$\,J/kg\,K \\
    Capillary constant, $\Gamma$ & $10^{-10}$\,m \\
    \hline
  \end{tabular}
\end{table}

\subsection{Observations of particle morphologies}

\begin{figure}[ht!]
    \centering
    \includegraphics[width=0.7\linewidth]{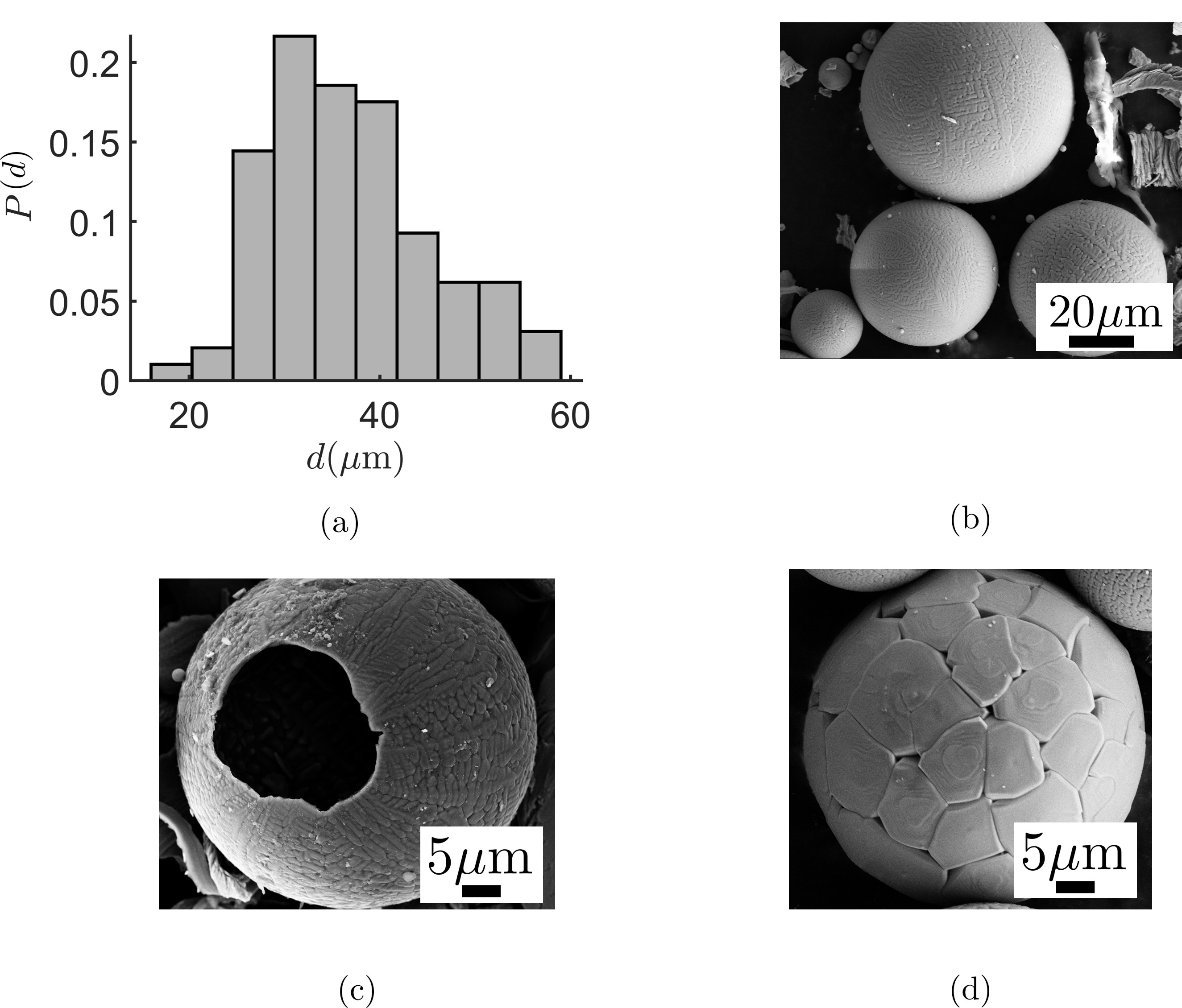}
    \caption{Scanning electron micrographs of particles obtained by abrasion: (a) Size distribution of particles; (b) Dendritic morphologies observed on spherical particles of various sizes; (c) A hollow particle with a thin outer shell; (d) Distinct cellular morphology on the particle surface.}
    \label{fig:shapesandmorphologies}
\end{figure}

Some of the primary features observed in the solidified metallic drops are reproduced in Fig.~\ref{fig:shapesandmorphologies}. Panel (a) shows the size distribution of the particles, with mean size of 36 $\mu$m. The scanning electron microscopy (SEM) images of individual particles are shown in Fig.~\ref{fig:shapesandmorphologies}(b). Spherical particles exhibit dendritic structures irrespective of their diameters, indicating that the conditions prevailing during formation permit morphology selection across the entire size range. Notably, Fig.~\ref{fig:shapesandmorphologies}(c) displays a hollow spherical particle, with a dendritic structure visible in the interior. The thin shell seen in this panel appears to suggest solidification of a bubble as opposed to a drop. Finally, panel (d) shows a single particle with a very different surface morphology, reminiscent of a football surface. 



Based on these observations, it appears that the final surface patterns can be  broadly classified into three categories, seemingly independent of particle size: (i) cellular (\emph{cf.} Fig.~\ref{fig:shapesandmorphologies}(d)), (ii) dendritic ({\emph{cf.} Figs.~\ref{fig:shapesandmorphologies}(b) and (c)), and (iii) mixed, where both cells and dendrites coexist. The specific morphology (or combination of morphologies) observed in a given particle is determined by the thermal and kinetic conditions prevailing during solidification. These are analyzed systematically in the next section.
\section{Analysis and Results}
\label{sec:analysisResults}

We present an analysis of the solidification problem by considering nucleation thermodynamics, growth kinetics and stability for two complementary solidification modes under confinement and in the presence of background curvature.

\subsection{Physical processes and underlying assumptions}
\label{subsec:physicalProcesses}

To establish a foundation for the subsequent analysis, we first evaluate the dominant heat transfer mechanisms during the early stages of solidification. Given the small size of particles produced via abrasion and atomisation, it is essential to assess whether a lumped heat capacitance model is appropriate for estimating initial cooling rates, until the onset of nucleation and growth. This is done by evaluating the Biot number ($Bi$), which compares internal conduction to convective heat transfer from the surface.

Under the experimental conditions described in Sec.~\ref{sec:experimental}---specifically, a wheel diameter of $170$\,mm and a rotation speed of $2800$\,rpm---the estimated particle ejection velocity is $v \sim 25$\,m/s. This velocity lies within the typical range reported for atomisation processes ($8$--$100$\,m/s)~\cite{bewlay1990modeling}, suggesting that the thermal transport considerations here are broadly applicable to both abrasion and gas atomisation.

To estimate $Bi$, we first determine the convective heat transfer coefficient $h$ using the Ranz--Marshall correlation~\cite{ranz1952evaporation}:
\begin{equation}
Nu = 2 + 0.6\,Re^{1/2}Pr^{1/3},
\end{equation}
where $Nu$ is the Nusselt number, and $Re$ and $Pr$ are the Reynolds and Prandtl numbers, respectively. Using standard thermophysical properties for air and helium~\cite{bergman2011introduction, hilsenrath1954viscosity}, we find that for a $50\,\mu$m particle travelling at $25$\,m/s in air, $Nu \simeq 6$, while for a particle at $100$\,m/s in helium, $Nu \simeq 10$. This indicates that convective heat removal is less efficient in abrasion than in gas atomisation.

For the present abrasion case, with $Nu = 6$, the corresponding heat transfer coefficient is $h = 3.6 \times 10^3$\,W/m$^2$K, yielding $Bi = 1.2 \times 10^{-3} $. This confirms that internal temperature gradients in the liquid drop can be safely neglected, and that a lumped thermal analysis is valid for estimating initial cooling rates, before the onset of nucleation. Once nucleation occurs, the drop is no longer isothermal; in fact it is the local temperature field that determines the final microstructural patterns.

These considerations justify the following assumptions that underpin our subsequent analysis:
\begin{enumerate}[label=(\roman*)]
\item Drop is initially uniformly undercooled:  Given that $Bi \ll 1$, the entire liquid drop is assumed to be initially at a uniformly undercooled temperature $T = T_m - \Delta T$. Further, we assume heterogeneous nucleation occurs at an equivalent undercooling temperature, since the undercooling required for homogeneous nucleation in Fe is approximately $295^\circ$C, which is highly unlikely under the conditions described in Sec.~\ref{sec:experimental}.
\item  Dual nucleation pathways:  Nucleation may occur both within the bulk and on the surface of the molten droplet, consistent with observations of competing growth fronts in spherical geometries.
\item Conduction-dominated heat removal:  Heat loss is assumed to occur primarily via conduction into the undercooled liquid and the partially solidified material. While convective losses at the boundary are neglected for analytical tractability, their effect is indirectly accounted for by having a range of undercooling $\Delta T$. This is motivated by the fact that the local temperature field during growth is modulated by the particle's residence time in air. Large residence time implies, more heat removal by convection, lower $T_\infty$, and, hence, larger $\Delta T$.
\end{enumerate}

These assumptions allow us to focus on the intrinsic thermodynamic and geometric factors governing solidification, without the added complexity of spatially varying boundary conditions. In the following sections, we build upon this framework to analyse nucleation thermodynamics, growth kinetics, and interface stability in spherical metallic particles.

\subsection{Two competing solidification modes: radial outward vs.\ circumferential growth}
\label{subsec:competingsolidificationmodes}

As established in Sec.~\ref{subsec:physicalProcesses}, the thermal conditions during particle flight favour rapid solidification, with nucleation likely occurring both within the bulk and on the surface of the molten droplet. To facilitate analytical treatment, we distinguish between two idealised but complementary solidification modes: \emph{radial outward} (RO) growth, initiated from a nucleus located at or near the centre of the droplet, and \emph{circumferential growth} (CG), initiated from a surface nucleation site and proceeding first along the outer curved surface, followed by inward growth (see Fig.~\ref{fig:nucleationSchematic}).

While RO growth is the conventional mode expected in spherical droplets~\cite{gill1981rapid}, the high surface undercooling and rapid heat removal characteristic of the present process make CG equally plausible~\cite{levi1982heat}. These two modes differ not only in their thermal and kinetic characteristics but also in their macroscopic consequences, particularly with respect to solidification shrinkage and defect formation.

To illustrate these differences, consider a spherical molten droplet of radius $a$, initially at temperature $T_0$. The total volumetric shrinkage upon solidification, $\Delta V = V_L - V_S$, can be estimated from mass conservation using the densities $\rho_L$ and $\rho_S$ of the liquid and solid phases:
\begin{equation}
\frac{\Delta V}{V_L} = 1 - \frac{\rho_L}{\rho_S}.
\end{equation}
For pure Fe (see Table~\ref{tbl:materialPropsFe}), this yields $\Delta V/V_L \simeq 0.1$.
\begin{figure}[ht!]
	\centering
	\includegraphics[width=0.9\textwidth]{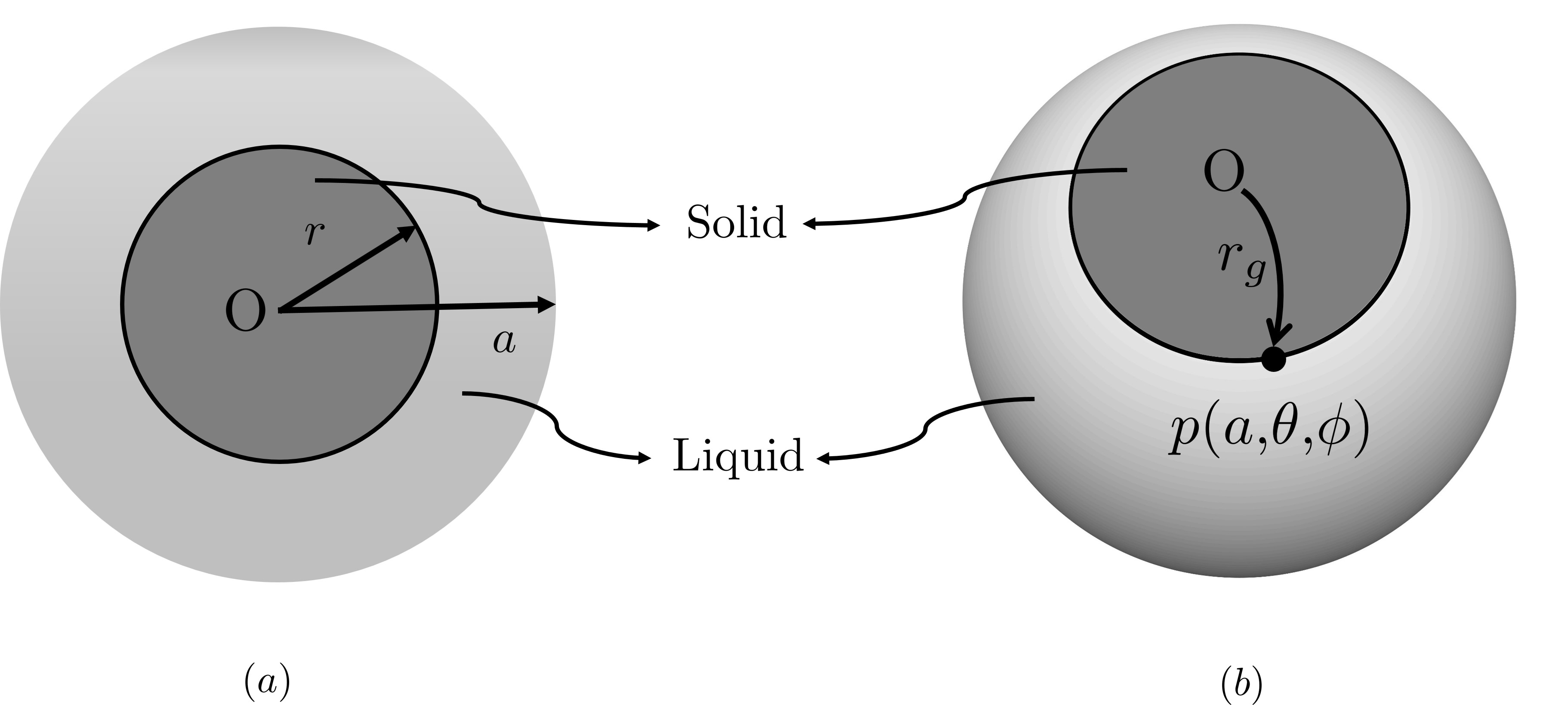}
	\caption{(a) Schematic showing a nucleus of radius $r$ growing radially outward inside a sphere of radius $a$. Panel (b) shows a nucleus of geodesic radius $r_g$ on the surface of a sphere of radius $a$. Coordinates of any point on the surface are specified by ($a,\theta,\phi$).}
	\label{fig:nucleationSchematic}
\end{figure}
The manifestation of this shrinkage depends critically on the dominant solidification mode. In the RO mode, the outer surface remains liquid during most of the solidification process, allowing the 10\% volume reduction to be accommodated by a decrease in the final particle radius. In contrast, CG mode initiates solidification at the surface, forming a rigid shell that constrains subsequent shrinkage. As the remaining liquid solidifies inward, the inability to contract externally leads to the formation of an internal void. For $\Delta V/V_L \simeq 0.1$, a $30\,\mu$m radius droplet would be expected to develop a central void of nearly $24\,\mu$m in diameter.

Thus, the presence of an internal void serves as strong morphological evidence for circumferential-dominated solidification.

\subsection{Nucleation thermodynamics of RO and CG modes}
\label{subsec:nucleationThermodynamics}

Building on the physical framework established in Sec.~\ref{subsec:physicalProcesses}, we now examine the thermodynamic conditions under which nucleation occurs in the two competing solidification modes discussed in Sec.~\ref{subsec:competingsolidificationmodes}: radial outward (RO) and circumferential growth (CG). Nucleation of the solid phase within an undercooled liquid is driven by the balance between the reduction in bulk free energy and the cost of creating a solid--liquid interface~\cite{kurz2023fundamentals}. For each mode, we evaluate the corresponding free energy change to determine the critical nucleus size required for spontaneous growth. Although the analysis is presented for homogeneous nucleation, the results also apply to heterogeneous nucleation, if the usual change in undercooling were made~\cite{kurz2023fundamentals}. 

\subsubsection{Radial outward (RO) mode} In the RO mode, a spherical solid nucleus of radius $r$ forms within the bulk of an isothermal, undercooled liquid droplet of size $a$, see schematic in Fig.~\ref{fig:nucleationSchematic}(a). The change in Gibbs free energy $G^{RO}$ is independent of the droplet radius $a$ and is given by:
\begin{equation}
\label{eqn:G_RO_mode}
G^{RO}(r) = 4\pi r^2\gamma - \dfrac{4\pi}{3}r^3f,
\end{equation}
where $f = L_v \Delta T / T_m$ is the bulk free energy change per unit volume, expressed in terms of the undercooling $\Delta T = T_m - T$, the drop's (assumed) uniform temperature $T$, melting temperature $T_m$, and latent heat per unit volume $L_v$; $\gamma$ is the solid--liquid surface tension. The free energy attains a maximum at the critical radius:
\begin{equation}
\label{eqn:rc_RO_mode}
\rcRO = \dfrac{2\gamma}{f} = \dfrac{2\gamma T_m}{L_v \Delta T},
\end{equation}
beyond which the solid phase becomes thermodynamically favoured and grows spontaneously.
\subsubsection{Circumferential growth (CG) mode} In the CG mode, nucleation occurs on the surface of the droplet, with the nucleus defined by a geodesic radius $r_g$ measured along the surface latitude, see schematic in Fig.~\ref{fig:nucleationSchematic}(b). Assuming a shell thickness $\delta$ in the radial direction, the corresponding free energy change $G^{CG}$ is:
\begin{equation}
\label{eqn:G_CG_mode}
G^{CG}(r_g,a,\delta) = 2\pi a \delta \left[ \gamma \sin\left(\frac{r_g}{a}\right) - \frac{a L_v \Delta T}{T_m} \left(1 - \cos\left(\frac{r_g}{a}\right) \right) \right],
\end{equation}
where $a$ is the droplet radius, the other symbols are the same as those in Eq.~\ref{eqn:rc_RO_mode}. The critical geodesic radius $\rcCG$ is obtained by maximising $G^{CG}$:
\begin{equation}
\label{eqn:rc_CG_mode}
\rcCG = a \tan^{-1}\left( \frac{1}{a} \frac{\gamma}{f} \right) = a \tan^{-1}\left( \frac{1}{a} \frac{\gamma T_m}{L_v \Delta T} \right).
\end{equation}
In the limit $\lim_{x \to 0} \tan^{-1}(x) \sim x$, this expression reduces to the classical result for a two-dimensional nucleus in planar geometry (see\suppMat{S.1}).
\subsubsection{Critical radii and free energies}

The variation of $\rcCG$ with undercooling $\Delta T$ is reproduced in Fig.~\ref{fig:criticalNucleus_CG} for different steel/Fe droplet sizes. Note that the horizontal axis is non-dimensionalized as $\Delta = \Delta T/T_m$ and the vertical axis represents the (dimensionless) critical radius $\rcCG L_v/\gamma$. Two trends are evident: (i) the effect of droplet size on $\rcCG$ is significant only at low undercooling, and (ii) for a fixed low $\Delta$, smaller droplets exhibit smaller critical radii, making surface nucleation more probable. Thus, smaller droplets with lower undercooling can, at first glance, appear to favour nucleation and growth via the CG mode.

\begin{figure}[ht!]
  \centering
  \includegraphics[width=0.8\textwidth]{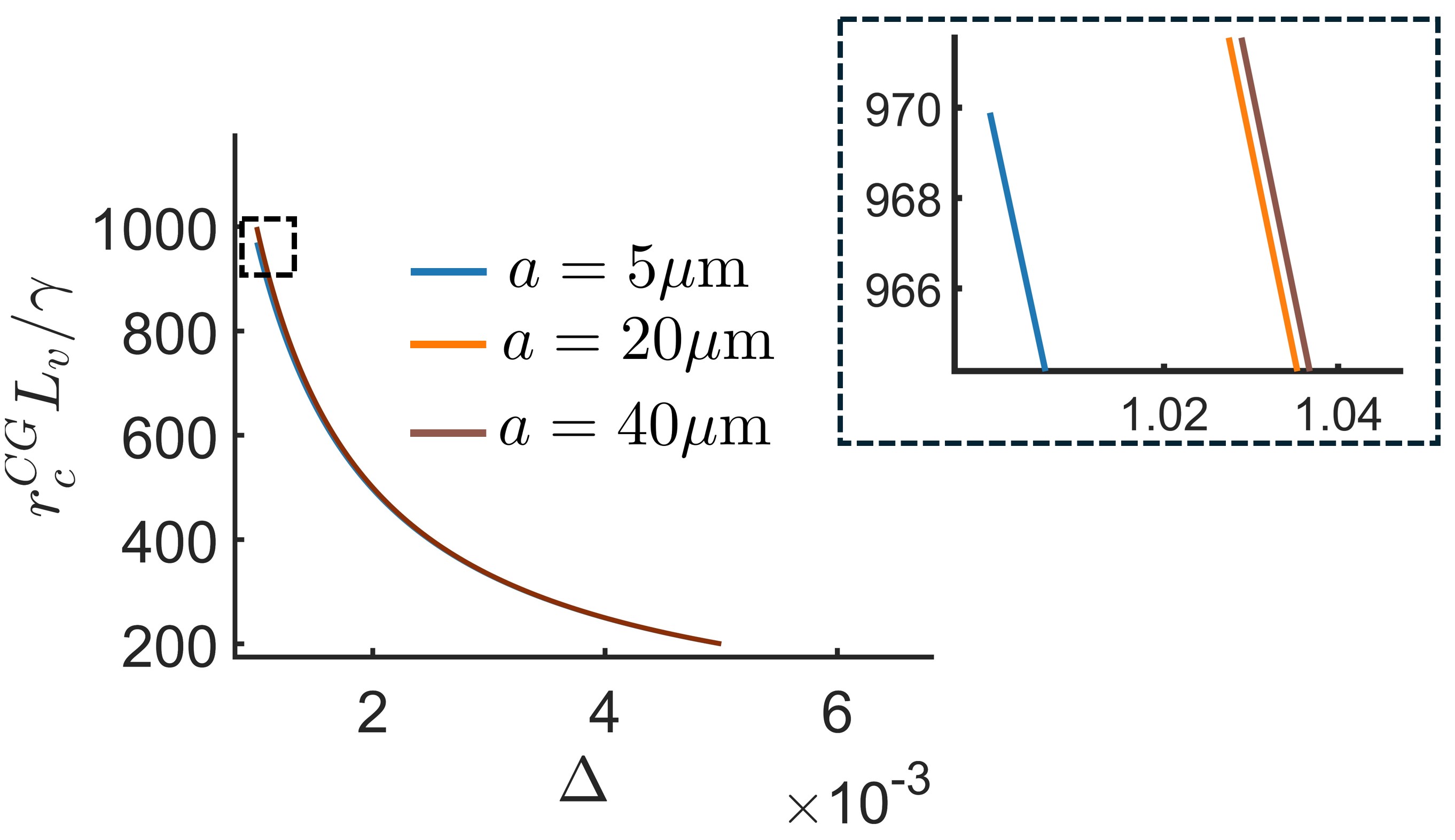}
  \caption{Variation of dimensionless critical radius of nucleation for the CG mode with undercooling $\Delta $ for various values of sphere radius $a$. Inset shows a magnified view of the critical radius at low undercooling $\Delta$.}
  \label{fig:criticalNucleus_CG}
\end{figure}

It is important to note that $\rcRO$ and $\rcCG$ represent different geometric dimensions (in the radial and azimuthal directions, respectively) and cannot be directly compared. To assess the relative likelihood of each mode, we compute the free energy changes $G^{RO}$ and $G^{CG}$ at their respective critical nucleus sizes using Eqs.~\ref{eqn:G_RO_mode}--\ref{eqn:rc_CG_mode}. Figure~\ref{fig:G_nucleation_RO_CG} shows the resulting free energy barriers as a function of normalised undercooling $\Delta$. Three key observations emerge. Firstly, the RO mode, being volumetric, shows no dependence on droplet size ( \mystar{Green4} curve ). Secondly, the CG mode exhibits weak radius dependence, as seen from the nearly overlapping curves for $a = 5$, $20$, and $40\,\mu$m. Consequently, and contrary to what the critical radius might suggest, the occurrence of CG modes is not a strong function of drop radius. Finally, the free energy curves intersect at a critical undercooling $\Delta_c$ (assuming $\delta = 0.5\,\mu$m), marked by dash-dot line in the figure. For $\Delta < \Delta_c$, the CG mode has a lower nucleation barrier and is energetically favoured; for $\Delta  > \Delta_c$, the RO mode becomes dominant. This cross-over is shown clearly in the inset to Fig.~\ref{fig:G_nucleation_RO_CG}.

\begin{figure}[ht!]
  \centering
  \includegraphics[width=0.8\textwidth]{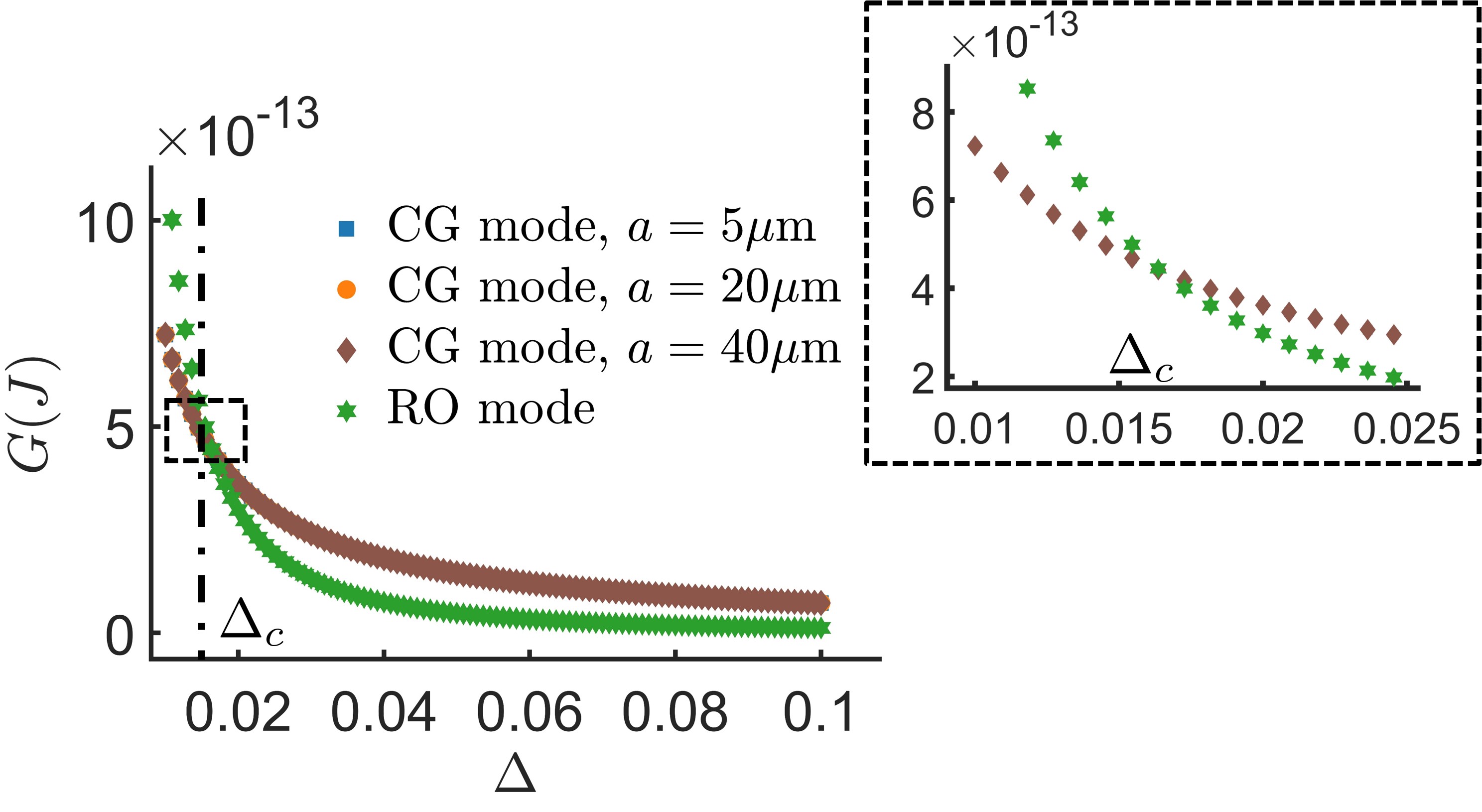}
  \caption{Gibbs free energy change (in Joules) for nucleation events via the RO and CG modes at their respective critical radii. For lower undercooling $\Delta$, the CG mode is energetically favoured. Inset shows a magnified view of the critical undercooling $\Delta_c$. }
  \label{fig:G_nucleation_RO_CG}
\end{figure}

In summary, the CG mode is more likely to dominate in droplets with lower undercooling. Hence, particles that solidify earlier---i.e., with lower $\Delta T$ due to shorter residence times in air---are more likely to exhibit voids, indicating circumferential growth, see Sec.~\ref{subsec:competingsolidificationmodes}.

\subsection{Growth and solidification times for RO and CG modes}
\label{subsec:growthKinetics}

Having established the thermodynamic conditions favouring either the radial outward (RO) or circumferential growth (CG) mode, we now analyse the kinetics of solidification. Under the experimental conditions described in Sec.~\ref{sec:experimental}, and given the small Biot number (Sec.~\ref{subsec:physicalProcesses}), heat conduction is the dominant mechanism governing the growth of the solid phase from an established nucleus. The governing equations consist of two parabolic partial differential equations (PDEs) for the temperature fields in the solid($T_S$) and liquid phases($T_L$):
\begin{equation}
\label{eqn:stefanPDE_nonDim}
\mathcal{S}\frac{\partial T_{L,S}}{\partial t}=\frac{1}{r^2}\dfrac{\partial}{\partial r}\bigg(r^2\frac{\partial T_{L,S}}{\partial r}\bigg)+ \frac{1}{r^2\sin\theta}\dfrac{\partial}{\partial \theta}\bigg(\sin\theta\frac{\partial T_{L,S}}{\partial \theta}\bigg)+\frac{1}{r^2\sin^2\theta}\dfrac{\partial^2 T_{L,S}}{\partial \phi^2}
\end{equation}
here $\mathcal{S}=\dfrac{C(T_m-T_\infty)}{L_v}$ is the Stefan number. Note that this equation is presented in dimensionless form, with $T_{L,S}$ being the difference between the instantaneous temperature in the liquid/solid phase and the ambient temperature $T_{\infty}$, scaled by $T_m - T_{\infty}$. The radial coordinate $r$ is non-dimensionalized by some length scale $\xi$ (different for RO and CG modes) and time $t$ correspondingly non-dimensionalized by $\mathcal{S} \alpha_L/a^2$. This procedure is described in more detail in \suppMat{S.2}.

These two PDEs are coupled by an energy balance relation at the solid-liquid interface, commonly referred to as the Stefan condition~\cite{rubinshteuin1971stefan} 
\begin{equation}
\left(\frac{k_S}{k_L}\nabla T_S- \nabla T_L\right)\biggl|_{r=\mathcal{R}}.\hat{n}=\frac{d \mathcal{R}}{dt} \label{eqn:stefanBC_nonDim}
\end{equation}
where the right hand side is the instantaneous interface velocity in the normal $\hat{n}$ direction, denoted by $\dfrac{d\mathcal{R}}{dt}$, where $\mathcal{R}$ is the location of the interface.

This classical Stefan problem is analytically tractable only in simplified geometries, such as infinite one-dimensional domains~\cite{alexiades2018mathematical}. For the present case, we consider two distinct Stefan problems corresponding to the RO and CG modes, each with its own geometry, boundary conditions and non-dimensionalization.

\subsubsection{Growth from a critical nucleus in the RO mode}

For the RO mode, we consider a solid nucleus of initial radius $\rcRO$ (Eq.~\ref{eqn:rc_RO_mode}) growing outward within a spherical liquid droplet of radius $a$. The goal is to determine the time evolution of the solid radius $\mathcal{R}=R(t)$ and estimate the total solidification time $t_s^{RO}$, defined by $R(t_s^{RO}) = 1$.

We restate the governing PDE (see \suppMat{S.2}) using $\xi = a$ as the characteristic length scale. Using the quasi-stationary approximation~\cite{gill1981rapid, alexiades2018mathematical}, which is valid when $\mathcal{S} \ll 1$, the time-dependent term is neglected. This condition holds when the maximum possible undercooling $\Delta = (T_m - T_\infty)/T_m \lesssim 0.3$ for Fe. For the experimental conditions discussed in this manuscript, we have $\Delta = 0.03$ (\emph{cf.} Sec.~\ref{subsec:nucleationThermodynamics}). A point worth noting is that the undercooling $\Delta$ required for growth is much lower than that of undercooling required for nucleation. Therefore, once the nucleation barrier is overcome, growth can occur at lower undercooling, albeit at a slower rate. 

The boundary conditions for interface growth in the RO problem are:
\begin{align}
  T_{L,S}(r=R,t) &= 1 - \dfrac{2\Gamma}{R\Delta}, \label{bc_ri} \\
  T_S(r=0,t) &\ne \infty, \label{bc_rzero} \\
  T_L(r=1,t) &= 0, \label{bc_router}
\end{align}
where $\mathcal{R} = R$ (interface location), $r$ (radial coordinate) and $\Gamma$ (capillary length) are all non-dimensionalized by $a$ (see \suppMat{S.2}).

The solutions for the temperature fields are given by:
\begin{equation}
\label{eqn:temperature_RO_mode}
T_S(r) = 1 - \dfrac{2\Gamma}{R\Delta}, \quad\quad T_L(r) = \left( \dfrac{R - 2\Gamma/\Delta}{r} \right) \left( \dfrac{r - 1}{R - 1} \right).
\end{equation}

The temperature inside the solid remains constant, increasing toward $T_m$ as the front approaches $R = 1$. Applying the Stefan condition at the interface $r = R(t)$, we obtain:
\begin{equation}
\label{eqn:Rt_RO_mode}
\frac{dR}{dt} = \dfrac{1}{R^2(1 - R)} \left\{ R - \dfrac{2\Gamma}{\Delta} \right\}.
\end{equation}

The interface position $R(t)$ is obtained by integrating Eq.~\ref{eqn:Rt_RO_mode} from the initial condition $R(0) = \rcRO/a$ to $R(t_s^{RO}) = 1$. Due to the singularity at $R = 1$, we define solidification completion as $|R - 1| \leq 0.01$\footnote{This is due to the quasi steady approximation, which is valid only for domains of infinite extent }. We also confirm that $dR/dt > 0$ for $R \geq \rcRO/a$, ensuring stability of the nucleus under the quasi-stationary approximation.

\subsubsection{Growth from a critical nucleus in the CG mode}
\label{subsec:growthinCGmode}

For the CG mode, we adopt $\xi = \pi a/2$ as the characteristic length scale and the radius of the initial solid to be $\mathcal{R}=R_g$, as measured along the geodesic radius (see Fig.~\ref{fig:nucleationSchematic}). The boundary conditions are:
\begin{align}
  T_{S,L}(r_g = R_g, t) &= 1 - \dfrac{\Gamma \kappa_g}{\Delta}, \label{bc_si} \\
  T_S(r_g = 0, t) &\ne \infty, \label{bc_szero} \\
  T_L(r_g = 1, t) &= 0, \label{bc_souter}
\end{align}
where $r_g$ is the geodesic coordinate non-dimensionalized by $\xi = \pi a/2$ and $\kappa_g = (\pi/2)\cot(\pi R_g/2)$ is the non-dimensional geodesic curvature. The equator is chosen as the outer boundary for the solidifying front on the spherical surface.

Under the same quasi-stationary approximation, the temperature fields are:
\begin{align}
  T_S(r_g) &= 1 - \dfrac{\Gamma \kappa_g}{\Delta}, \label{sssols} \\
  T_L(r_g) &= \dfrac{1 - \Gamma \kappa_g/\Delta}{\log\left[ \tan\left( \dfrac{\pi R_g}{4} \right) \right]} \log\left[ \tan\left( \dfrac{\pi r_g}{4} \right) \right]. \label{ssliqs}
\end{align}

As in the RO case, the solid temperature remains constant. Applying the Stefan condition at $r_g = R_g(t)$, we obtain:
\begin{equation}
\label{eqn:Rt_CG_mode}
\frac{dR_g}{dt} = \dfrac{-2}{\pi \sin(\pi R_g/2)} \left( \dfrac{1 - \Gamma \kappa_g/\Delta}{\log\left[ \tan(\pi R_g/4) \right]} \right).
\end{equation}

Integrating from $R_g(0) = \rcCG$ to $R_g(t_s^{CG}) = 1$ yields the solidification time $t_s^{CG}$. As before, we define completion as $|R_g - 1| \leq 0.01$ to avoid singular behaviour.

\subsubsection{Comparison of solidification times}

The nucleation analysis in Sec.~\ref{subsec:nucleationThermodynamics} showed that both RO and CG modes may initiate depending on droplet radius $a$ and normalised undercooling $\Delta$. The corresponding solidification times $t_s^{RO}$ and $t_s^{CG}$ are indicative of the relative growth rates of each mode.

\begin{figure}[ht!]
  \centering
  \includegraphics[width=1\textwidth]{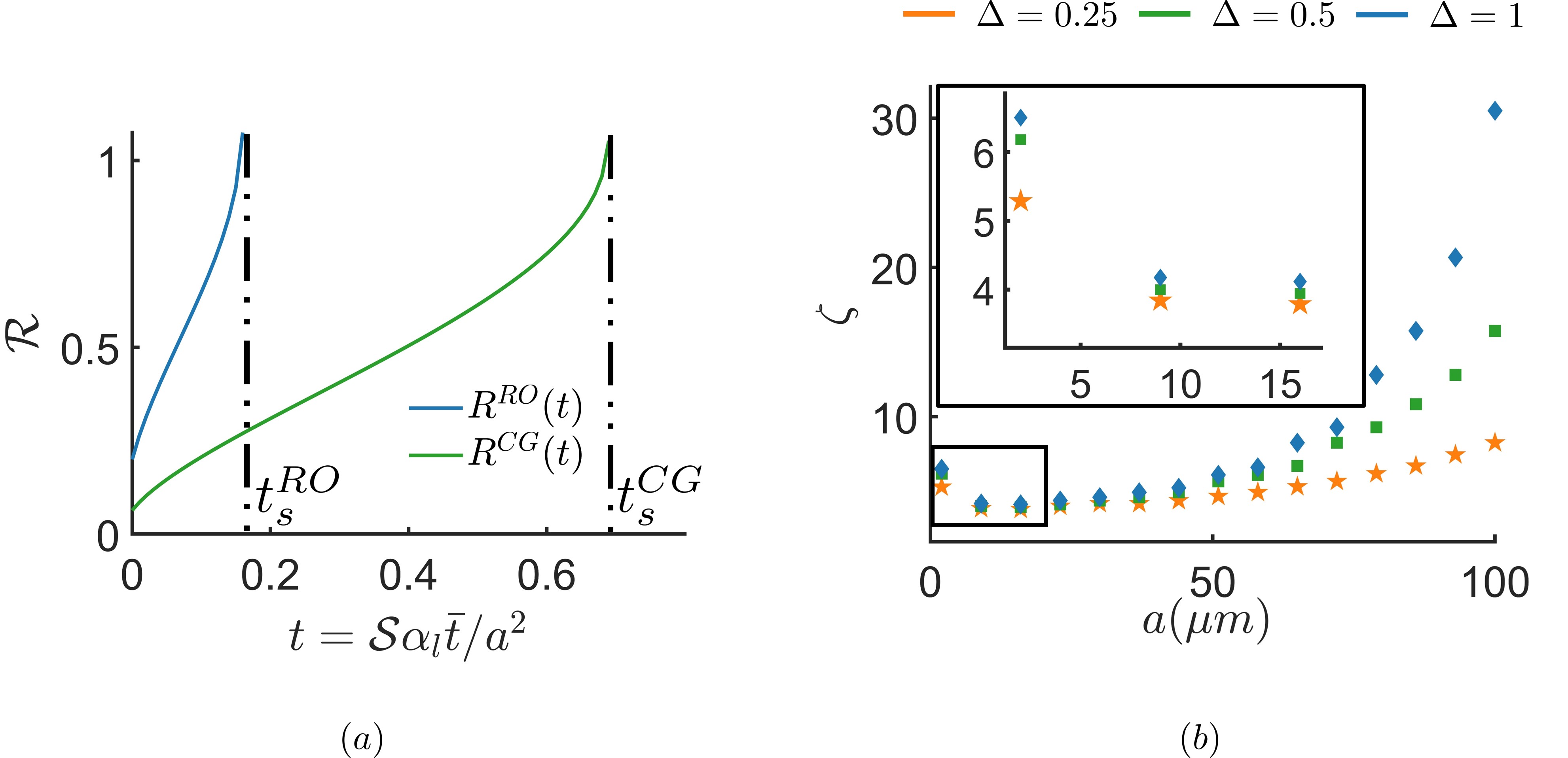}
  \caption{(a) Time evolution of non-dimensionalised growth fronts $\mathcal{R}(t)$ in RO and CG modes. (b) Variation of the ratio of solidification times ($\zeta = t_s^{CG}/t_s^{RO}$) with sphere radius $a$ for various values of $\Delta \times 10^3$. Inset shows the dependence of $\zeta$ on $a$ for very small $a \sim 10\,\mu$m.}
  \label{fig:growth_Rt_tc}
\end{figure}

Figure~\ref{fig:growth_Rt_tc}(a) shows the temporal evolution of the non-dimensional interface positions $\mathcal{R}(t)$ for both RO and CG modes, computed at a fixed undercooling $\Delta= 0.03$ and droplet radius $a = 5\,\mu$m. The curves for the RO mode ($R(t)$, blue) and CG mode ($R_g(t)$, green) originate from their respective critical nucleus sizes $\rcRO$ and $\rcCG$, and evolve toward the outer boundary. The time at which each curve reaches unity corresponds to the solidification time, denoted by $t_s^{RO}$ and $t_s^{CG}$, for RO and CG modes, respectively.

To quantify the relative growth rates, we define the ratio $\zeta = t_s^{CG}/t_s^{RO}$: when $\zeta \gg 1$, the RO mode outpaces CG growth; and vice-versa when $\zeta \ll 1$. Figure~\ref{fig:growth_Rt_tc}(b) presents the variation of $\zeta$ with droplet radius $a$ for several values of undercooling $\Delta$. Two key observations can be made : (i) for the entire range of $a$ and $\Delta$ considered, $\zeta > 1$, indicating that RO growth is generally faster; and (ii) for smaller droplets ($a < 50\,\mu$m), the solidification times are comparable ($\zeta = \mathcal{O}(1)$), suggesting that neither mode is strongly dominant in this regime. The inset shows that undercooling has minimal influence on $\zeta$ for small $a$.

As the drop radius $a$ increases beyond $50\,\mu$m, the effect of undercooling becomes more pronounced. Higher undercooling accelerates RO growth relative to CG, resulting in $\zeta \gg 1$. Thus, for larger droplets at high undercooling, solidification is expected to proceed predominantly via the RO mode. Importantly, even when CG nucleation is thermodynamically favoured (as discussed in Sec.~\ref{subsec:nucleationThermodynamics}), its growth may be overtaken by a simultaneously advancing RO front, particularly in smaller droplets. Consequently, morphological signatures of CG-mode solidification---such as interior voids---may not always be evident in post-mortem analyses. We revisit this important point in Sec.~\ref{sec:discussion}.

\subsection{Stability of growing RO and CG fronts}
\label{subsec:stabilityGrowthFronts}

Once nucleated, solidification fronts may exhibit a range of morphological instabilities during growth. Classical linear stability analysis---termed the Mullins-Sekerka (MS) analysis \cite{mullins1963morphological}---of planar fronts reveals the occurrence of cellular patterns due to preferential unstable growth of perturbations. We now perform analogous linear stability analysis of growth fronts in the RO and CG modes, accounting for curved geometries and finite domains.

The central idea is illustrated in Fig.~\ref{fig:schematic_perturbation}, where a growing interface is perturbed by a small-amplitude fluctuation $\epsilon$. The evolution of this perturbation determines the stability of the interface: if $\dot{\epsilon}/\epsilon > 0$, the perturbation grows and the interface is unstable; if $\dot{\epsilon}/\epsilon < 0$, the interface remains stable. Given the linear nature of the equations, we express the perturbations in terms of spherical harmonics (Fig.~\ref{fig:schematic_perturbation}(a)) for RO and cosine eigenfunctions for the CG mode (Fig.~\ref{fig:schematic_perturbation}(b)). 

\begin{figure}[ht!]
  \centering
  \includegraphics[width=0.8\textwidth]{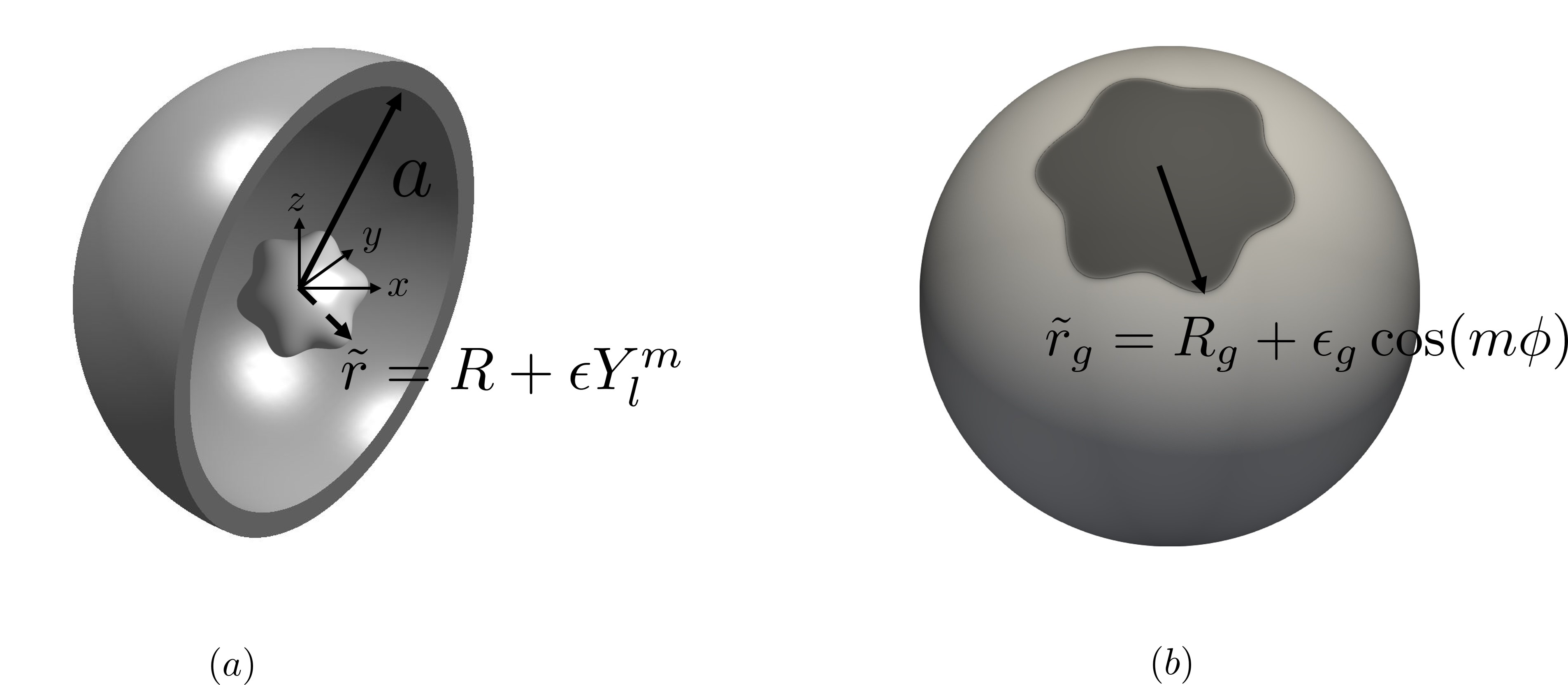}
  \caption{Schematic showing perturbed growing interfaces for the RO mode (a) and the CG mode (b). Perturbation amplitudes are exaggerated for clarity.}
  \label{fig:schematic_perturbation}
\end{figure}

\subsubsection{Linear stability analysis of RO growth mode}
\label{subsubsec:stability_RO}

Consider the radially growing solid nucleus described in Sec.~\ref{subsec:growthKinetics} with the interface perturbed as:
\begin{equation}
\tilde{r}(\theta, \phi, t) = R(t) + \epsilon(t) Y_l^m(\theta,\phi), \label{pertint}
\end{equation}
where $R$ is the interface location (dimensionless) as in the previous section, and $\epsilon=\bar{\epsilon}/a$ is the dimensionless perturbation amplitude. The presence of the perturbation only changes the Stefan boundary condition, which is now applied at the perturbed interface.

The boundary conditions are:
\begin{align}
T_{L,S}(r = \tilde{r}, t) &= 1 - \dfrac{\Gamma \kappa^p}{\Delta}, \label{bc_rpi} \\
T_S(r = 0, t) &\ne \infty, \label{bc_rpzero} \\
T_L(r = 1, t) &= 0, \label{bc_rpouter}
\end{align}
where $\kappa^p$ is the mean curvature of the perturbed interface. To linear order in $\epsilon$, it is given by:
\begin{equation}
\kappa^p = \dfrac{2}{R} + \left( \dfrac{l(l+1) - 2}{R^2} \right) \epsilon Y_l^m + \mathcal{O}(\epsilon^2).
\end{equation}

The unperturbed (or \lq base state\rq ) temperature fields in the solid ($T_S^b$) and liquid ($T_L^b$) are given by Eq.~\ref{eqn:temperature_RO_mode}. Perturbations to the interface induce corresponding perturbations in the temperature fields, which we expand as:
\[
T_{L,S}^p =
\begin{cases}
A \epsilon r^l Y_l^m, & 0 < r \le R, \\
\epsilon \left( \dfrac{B}{r^{l+1}} + C r^l \right) Y_l^m, & R \le r \le 1,
\end{cases}
\]
where $A$, $B$, and $C$ are constants determined from the boundary conditions in Eq.~\ref{bc_rpi} and $T_{L,S}^p(r = 1) = 0$.

The resulting perturbed temperature fields (to linear order in $\epsilon$) are:
\begin{equation}
T_S = T_S^b - \left[ \dfrac{\Gamma}{R^{l+2} \Delta } (l - 1)(l + 2) \right] r^l \epsilon Y_l^m + \mathcal{O}(\epsilon^2), \label{pertsol}
\end{equation}
\begin{equation}
T_L = T_L^b + \dfrac{1}{R^2} \left[ \dfrac{R - 2\Gamma/\Delta}{1 - R} - \dfrac{\Gamma}{\Delta}(l - 1)(l + 2) \right] \left[ \dfrac{r^{2l+1} - 1}{R^{2l+1} - 1} \right] \left( \dfrac{R}{r} \right)^{l+1} \epsilon Y_l^m + \mathcal{O}(\epsilon^2). \label{pertliq}
\end{equation}

Substituting Eqs.~\ref{pertsol} and \ref{pertliq} into the Stefan condition and equating coefficients of $Y_l^m$ yields the dispersion relation for perturbation growth:
\begin{align}
\label{growthrateRadial}
  \frac{\dot{\epsilon}}{\epsilon} &= \dfrac{k_L T_m (l - 1)}{L_v \bar{R}^2} \left\{ \Delta \left[ \dfrac{(l - 1) + \eta(l + 2)}{(\beta - 1)(\eta - 1)(l - 1)} \right] \right.\\
  \notag
&\quad - \left. \dfrac{\xi}{\beta} \left[ l(l + 2) \dfrac{k_S}{k_L} + \dfrac{2}{(\eta - 1)(\beta - 1)} + (l + 2)(l + 1) \left( \dfrac{1}{1 - \eta} - \dfrac{\eta}{1 - \eta} \left( \dfrac{2 - l(l - 1)(\beta - 1)}{(\beta - 1)(l^2 - 1)} \right) \right) \right] \right\}, 
\end{align}
where $\beta = R$ (non-dimensional), $\eta = R^{2l+1}$, and $\xi = \Gamma$.

In the limit $\beta \to 0$ (i.e., $a \to \infty$), this relation reduces to the classical Mullins-Sekerka (MS) result~\cite{mullins1963morphological}, as is to be expected. To better express the physical content in the dispersion relation, we rewrite it as:
\begin{equation}
\frac{\dot{\epsilon}}{\epsilon} = \dfrac{k_L T_m (l - 1)}{L_v \bar{R}^2} \left( \Pi_D^F - \Pi_S^F \right),
\end{equation}
where $\Pi_D^F$ and $\Pi_S^F$ represent the destabilising and stabilising contributions, respectively, in a finite domain (superscript $F$). The destabilising term $\Pi_D^F$ can be expressed in terms of its infinite-domain limit (superscript $I$) $\Pi_D^I = \Delta$:
\begin{align}
\notag \Pi_D^F &= \Pi_D^I \left[ \dfrac{(l - 1) + \eta(l + 2)}{(\beta - 1)(\eta - 1)(l - 1)} \right] \\
&= \Pi_D^I \left[ 1 + \beta + \beta^2 + \mathcal{O}(\beta^3) + (1 + p)(\eta + \beta \eta + \eta \beta^2) \right], \quad p = \dfrac{l + 2}{l - 1}. \label{destabforceseries}
\end{align}

Thus, the destabilising force is modified by the finite geometry, with the lowest-order term corresponding to the classical infinite-domain result. Unlike the planar case, the magnitude of $\Pi_D^F$ can be tuned by varying either the undercooling $\Delta$ or the interface position $\beta$.
\subsubsection{Linear stability analysis of CG growth mode}
\label{subsubsec:stability_CG}

We now analyse the stability of a solidification front growing along the surface of a sphere, as illustrated in Fig.~\ref{fig:schematic_perturbation}(b). The interface is described in terms of the geodesic radius $r_g$, with the unperturbed front located at $r_g = R_g$ (again, dimensionless, see Sec.~\ref{subsec:growthKinetics}). A sinusoidal perturbation is introduced as:
\begin{equation}
\tilde{r}_g(\phi, t) = R_g(t) + \epsilon_g(t) \cos(m\phi),
\end{equation}
where $\phi$ is the azimuthal angle and $\epsilon_g$ is the dimensionless perturbation amplitude. The non-dimensionalisation follows the same scaling as in Sec.~\ref{subsec:growthinCGmode}. To avoid singularities at $r = \pi a$, we restrict the domain to $\theta_0 < \theta < \pi/2$.

The boundary conditions are applied at the perturbed interface $r_g = \tilde{r}_g$:
\begin{align}
T_{L,S}(r_g = \tilde{r}_g, t) &= 1 - \dfrac{\Gamma \kappa_g^p}{\Delta}, \label{bc_spi} \\
T_S(r_g = 0, t) &\ne \infty, \label{bc_spzero} \\
T_L(r_g = 1, t) &= 0, \label{bc_spouter}
\end{align}
where $\kappa_g^p$ is the geodesic curvature of the perturbed front. Linearising in $\epsilon_g$, we obtain:
\begin{equation}
\kappa_g^p = \left( \dfrac{\pi}{2} \right) \cot\left( \dfrac{\pi R_g}{2} \right) - \dfrac{\pi/2}{\sin^2(\pi R_g/2)} \left[ 1 + \dfrac{\partial^2}{\partial \phi^2} \right] \epsilon_g \cos(m\phi) + \mathcal{O}(\epsilon_g^2).
\end{equation}

The unperturbed (or base state) temperature fields are given by Eqs.~\ref{sssols} and \ref{ssliqs}; the perturbed fields are expanded as:
\[
T_{L,S}^p =
\begin{cases}
C_0 \epsilon_g \cos(m\phi), & 0 < r_g \le R_g, \\
\left( C_1 \cosh\left\{ m \log\left[ \cot(r_g/2a) \right] \right\} + C_2 \sinh\left\{ m \log\left[ \cot(r_g/2a) \right] \right\} \right) \epsilon_g \cos(m\phi), & R_g \le r_g \le 1,
\end{cases}
\]
where, as in the RO case, the constants $C_0$, $C_1$, and $C_2$ are determined from boundary conditions. The final temperatures, to linear order in $\epsilon_g$, are:
\begin{align}
T_S &= T_S^b + \left( \dfrac{(\pi^2/4)\Gamma(1 - m^2)}{\Delta \sin^2(\pi R_g/2)} \right) \epsilon_g \cos(m\phi) + \mathcal{O}(\epsilon_g^2), \label{pertssols} \\
T_L &= T_L^b + \left( \dfrac{\pi C_2}{2} \right) \sinh\left\{ m \log\left[ \cot(\pi r_g/4) \right] \right\} \epsilon_g \cos(m\phi) + \mathcal{O}(\epsilon_g^2), \label{pertsliqs}
\end{align}
where
\begin{equation*}
C_2 = \left( \dfrac{ \dfrac{(\pi/2)\Gamma (1 - m^2)}{\Delta \sin^2(\pi R_g/2)} - \dfrac{(1 - \Gamma \kappa_g/\Delta)}{\log(\tan(\pi R_g/4)) \sin(\pi R_g/2)} }{ \sinh\left\{ m \log\left[ \cot(\pi R_g/4) \right] \right\} } \right).
\end{equation*}

Applying the Stefan condition yields the growth rate/ dispersion law:
\begin{multline}
\label{growthsurface}
\frac{\dot{\epsilon}_g}{\epsilon_g} = \frac{k_L T_m}{L_v a^2 \sin^2(\pi \beta_g/2)} \bigg[ \left( \dfrac{(\pi/2)\Gamma m(1 - m^2)}{\sin(\pi \beta_g/2)} - \dfrac{m(\Delta - \Gamma \kappa_g)}{\log(\tan(\pi \beta_g/4))} \right) \\
\times \coth\left\{ m \log\left[ \cot(\pi \beta_g/4) \right] \right\} + \dfrac{(\Delta - \Gamma \kappa_g)}{\log(\tan(\pi \beta_g/4))} \cos(\pi \beta_g/2) \bigg],
\end{multline}
where, analogous to the RO case, we have used $\beta_g = R_g$ (dimensionless) for comparison.
\begin{figure}[ht!]
	\centering
	\includegraphics[width=1\linewidth]{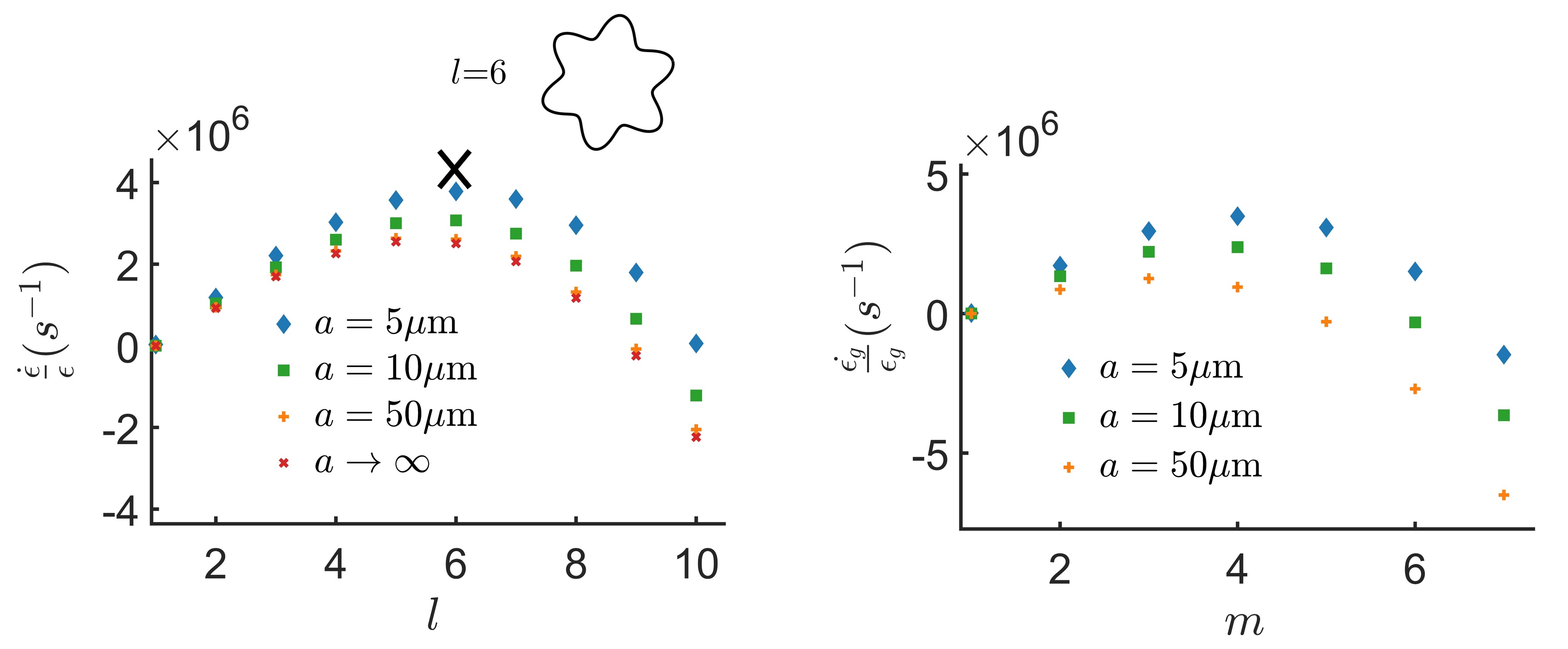}
	\caption{(a) Dispersion curves for various locations of outer boundary for radial growth. Inset shows a section of the perturbed sphere for $l_{max}=6$ (See ($\times$) marker) (b)  Dispersion curves for various locations of outer boundary for circumferential growth. }
	\label{fig:dispersion_combined}
\end{figure}

To understand the influence of $\beta,$ $\beta_g$ and $\Delta$ on interface stability, we evaluate the dispersion relation for various droplet sizes $a$, fixing the initial interface location at $\bar{R} = a \times R = 1\,\mu$m and $\bar{R}_g = \pi a/2\times R_g = 0.5\mu$m for the RO and CG modes, respectively. The stability/ dispersion curves for the RO mode are reproduced in Fig.~\ref{fig:dispersion_combined}(a). It is clear that as $a$ increases, the dispersion curves approach the infinite-domain limit, consistent with classical MS theory. Moreover, the peak growth rate decreases with increasing $a$, indicating slower perturbation growth in larger drops. One hence expects a wide range of unstable wavelengths to be prevalent for large $a$.

Figure \ref{fig:dispersion_combined}(b) shows the dispersion curves for the CG front, similar to the RO mode result. It is evident that the maximum growth rate reduces with increasing $a$, again implying that a wider variety of wavelengths may be operative in larger drops.

The fastest growing wavelength in these analyses (corresponding to highest $\dot{\epsilon}/\epsilon$) naturally introduces a microstructural length scale $\lambda$ into the present problem. This scale is governed by a competition between $\Delta$ and $\beta$ (or $\beta_g$). As $\beta$ decreases, higher-order contributions to the destabilising force diminish, and the destabilisation is driven primarily by $\Delta$ (see Eq.~\ref{destabforceseries}). Therefore, both $\beta$ ($\beta_g$) and $\Delta$ serve as controlling parameters for interface stability in the RO (CG) mode.

	

It is important to emphasise that the present linear stability analysis only captures the onset of instability. While it predicts the most unstable modes and their growth rates, it does not describe the nonlinear evolution of patterns such as cells or dendrites. A full nonlinear analysis would be required for quantitative predictions of such morphologies, which is beyond the scope of this study.

We conclude this section by discussing the implications of  growing unstable interface(s) on the overall morphology of the spherical particles. We examine two cases as before, an unstable radial front and an unstable front growing on the surface. In the event of the radial front reaching $\bar{r}=a$ faster than the surface front reaching $\bar{r}_g=\pi a/2$, the perturbations on the interface now, will act as obstacles to the front growing on the surface. Thus, the arms of the cells or dendrites growing on the surface will have to navigate through a distribution of obstacles in order to grow, depending on the nucleation rate and growth conditions. On the other hand if the perturbed front on the surface, covers most of the surface before the radial front reaches $\bar{r}=a$, the radial front will see a partly solidified shell with regions of undercooled liquid within the shell to grow into. Practically, this is the condition under which the solidified drop is most likely to show internal voids. 


\section{Interpretation of microstructure and morphologies}
\label{sec:Interpretation}

A brief summary of the analyses in Sec.~\ref{sec:analysisResults} is as follows. Firstly, simple mass balance predicted shrinkage-induced voids as large as $24\,\mu$m for a drop of radius $30\,\mu$m, suggesting that hollow particles may form due to volumetric contraction during solidification. Based on this, two complementary nucleation and growth modes---termed radial outward growth (RO) and circumferential growth (CG)---were identified. Nucleation thermodynamics (Sec.~\ref{subsec:nucleationThermodynamics}) outlined drop sizes $a$ and initial undercooling $\Delta$ values that favoured either mode. Growth kinetics in Sec.~\ref{subsec:growthKinetics}, under the quasi-stationary approximation, revealed that the RO mode dominates for larger droplets at high undercooling, based on the ratio of solidification times. Finally, linear stability analysis (Sec.~\ref{subsec:stabilityGrowthFronts}) provided conditions for the onset of morphological instability, and determined a characteristic microstructural length scale $\lambda$. Unlike classical Mullins--Sekerka theory, our analysis introduced the size ratios $\beta$ and $\beta_g$ as additional destabilising parameters alongside $\Delta$. Larger droplets were predicted to exhibit coarser microstructures at constant undercooling, and droplets of identical size could show different $\lambda$ depending on $\Delta$.

We now correlate these theoretical predictions with experimental observations of patterns on powder particle surfaces, \emph{cf.} Sec.~\ref{sec:experimental}. 

\begin{figure}[ht!]
  \centering
  \includegraphics[width=0.7\linewidth]{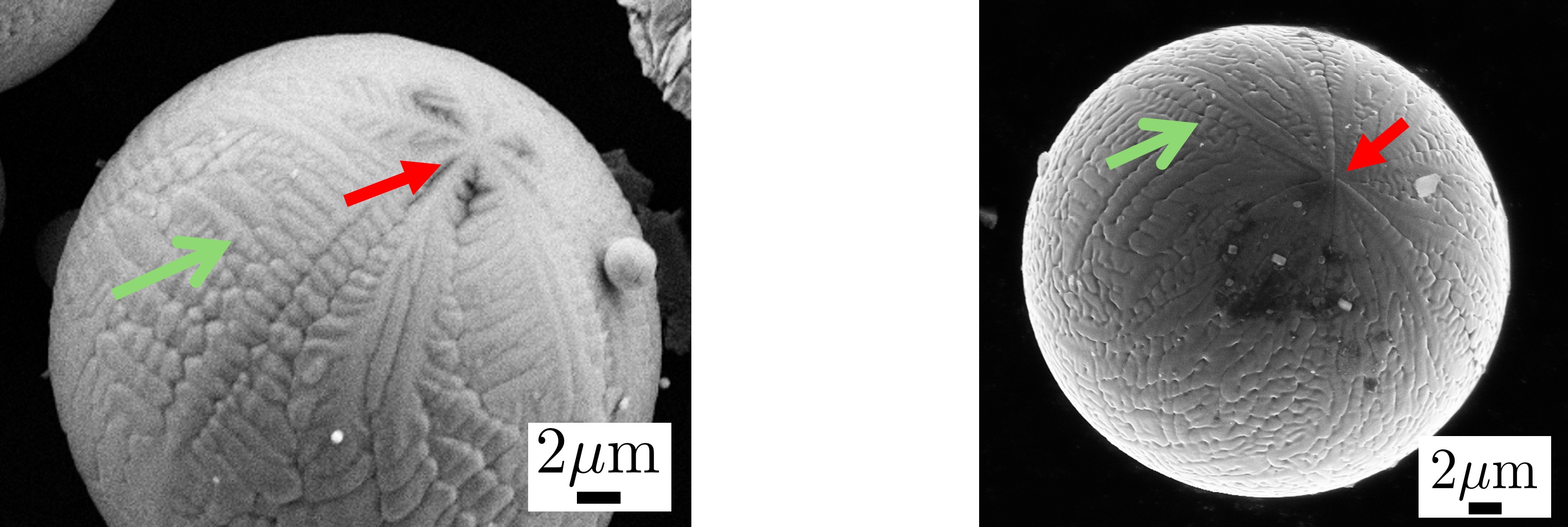}
  \caption{Nucleation on the surface of spherical particles. Dendrites emanating from surface nucleation sites are marked with red arrows; competing growth fronts are marked with green arrows.}
  \label{fig:nucleationexpcombined}
\end{figure}

Two typical solidified spheres with $a\sim 15\,\mu$m are reproduced in Fig.~\ref{fig:nucleationexpcombined}, showing dendrites emanating from surface nucleation sites (see red arrows). Initially, as the particles leave the abrasion zone (\emph{cf.} Sec.~\ref{sec:experimental}), the available undercooling $\Delta$ is low, placing all particles---regardless of size---left of the critical undercooling $\Delta_c$ in Fig.~\ref{fig:G_nucleation_RO_CG}, see Sec.~\ref{subsec:nucleationThermodynamics}. Thus, while nucleation is expected to begin preferentially on the surface, the inherent stochastic nature of the process allows for both internal nucleation as well as surface nucleation.

As the drops continue to cool via convection during their flight in air (or inert gas, in the case of atomization), $\Delta$ increases beyond $\Delta_c$, making bulk nucleation more favourable and allowing multiple growth fronts to coexist. Although surface nucleation likely initiates first, the RO front grows faster for any given $\Delta$ and $a$ (see Fig.~\ref{fig:growth_Rt_tc}(b), Sec.~\ref{subsec:growthKinetics}). Consequently, the radial front overtakes the slower CG front. This competition between the RO and CG modes may be seen in Fig.~\ref{fig:nucleationexpcombined} (see green arrows), where dendritic features on the surface trap pockets of undercooled liquid. These pockets are accessible to the RO front, resulting in equiaxed structures embedded within dendritic morphologies.

\begin{figure}[ht!]
	\centering
	\includegraphics[width=1\linewidth]{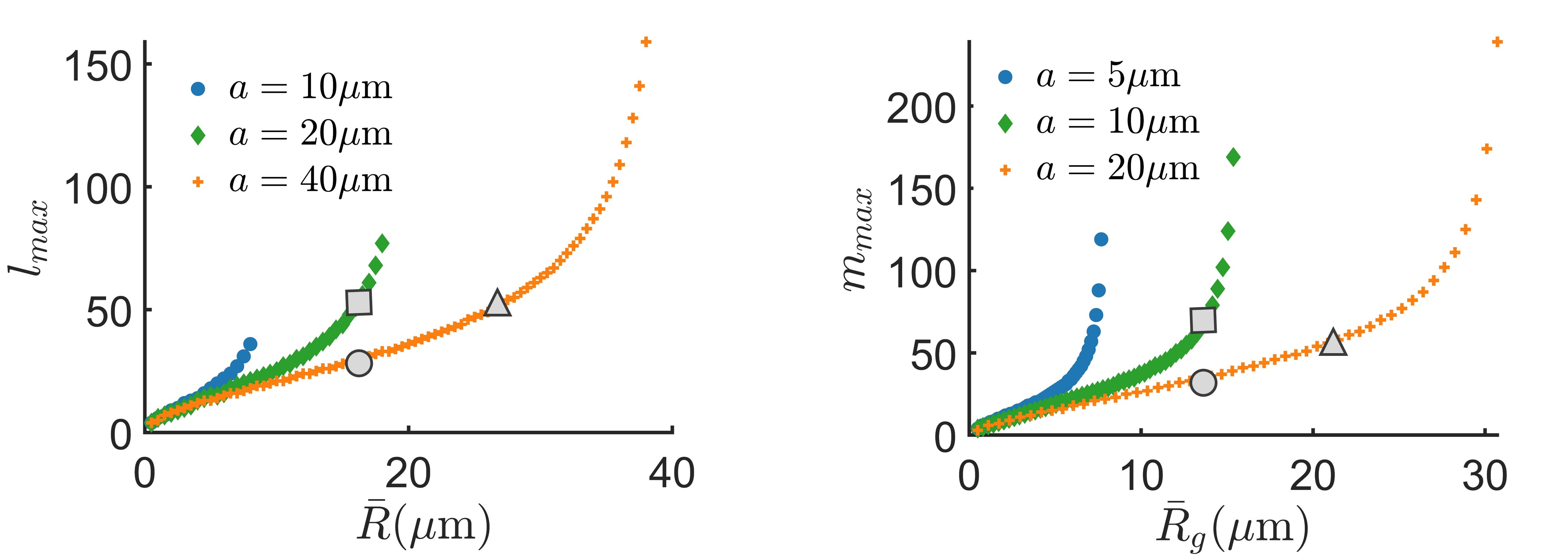}
	\caption{(a) Variation of maximum growth mode $l_{\text{max}}$ with initial seed size $\bar{R}$.(b) Variation of maximum growth mode $m_{\text{max}}$ with initial seed size $\bar{R}_g$.}
	\label{fig:dispandmmaxsurfacecombined}
      \end{figure}
  
The length-scale $\lambda$, and hence morphology actually observed, depends on the fastest growing wavelength, which is a function of $\Delta$ and $\beta$($\beta_g$) as obtained from Eq.~\ref{growthrateRadial} (\ref{growthsurface}). Effect of $\Delta$ on the microstructure is well understood in conventional solidification---increase in $\Delta$ will lead to microstructure refinement. On the other hand, $\beta$ (or $\beta_g$) is a new parameter that emerges from our analysis. Change in the fastest growing wavelength, denoted $l_{max}$ ($m_{max}$) for the RO (CG) mode, as a function of the initial seed size ($\beta$ or $\beta_g$ in Sec.~\ref{subsec:stabilityGrowthFronts}), and for various drop sizes $a$, are reproduced in Fig.~\ref{fig:dispandmmaxsurfacecombined}.

These graphs are deduced from the linear stability analysis presented in Sec.~\ref{subsec:stabilityGrowthFronts}, and are interpreted as follows. The parameter $\beta(\beta_g)$  can be varied either by varying $a$ or the interface location in dimensional units $\bar{R}(\bar{R}_g)$. For a fixed $\bar{R}(\bar{R}_g)$, corresponding to \mycircle{black, fill=gray!50} and \mysquare{black, fill=gray!50} markers in Fig. \ref{fig:dispandmmaxsurfacecombined}, a reduction in $a$ leads to an increase in $l_{\text{max}}$($m_{\text{max}}$). This implies a refinement in microstructure, since $\lambda \propto 1/l_{\text{max}}(m_{\text{max}})$. Conversely, for fixed $a$, corresponding to \mycircle{black, fill=gray!50} and \mytriangle{black, fill=gray!50} markers in Fig.~\ref{fig:dispandmmaxsurfacecombined}(a) and (b), increasing $\bar{R}(\bar{R}_g)$ also increases  $l_{\text{max}}$($m_{\text{max}}$) and hence reduces $\lambda$, leading to finer microstructures. Therefore, independent variation of $a$ and  $\bar{R}(\bar{R}_g)$ lead to consistent variation in microstructure. 
      

      \begin{figure}[ht!]
	\centering
	\includegraphics[width=1\linewidth]{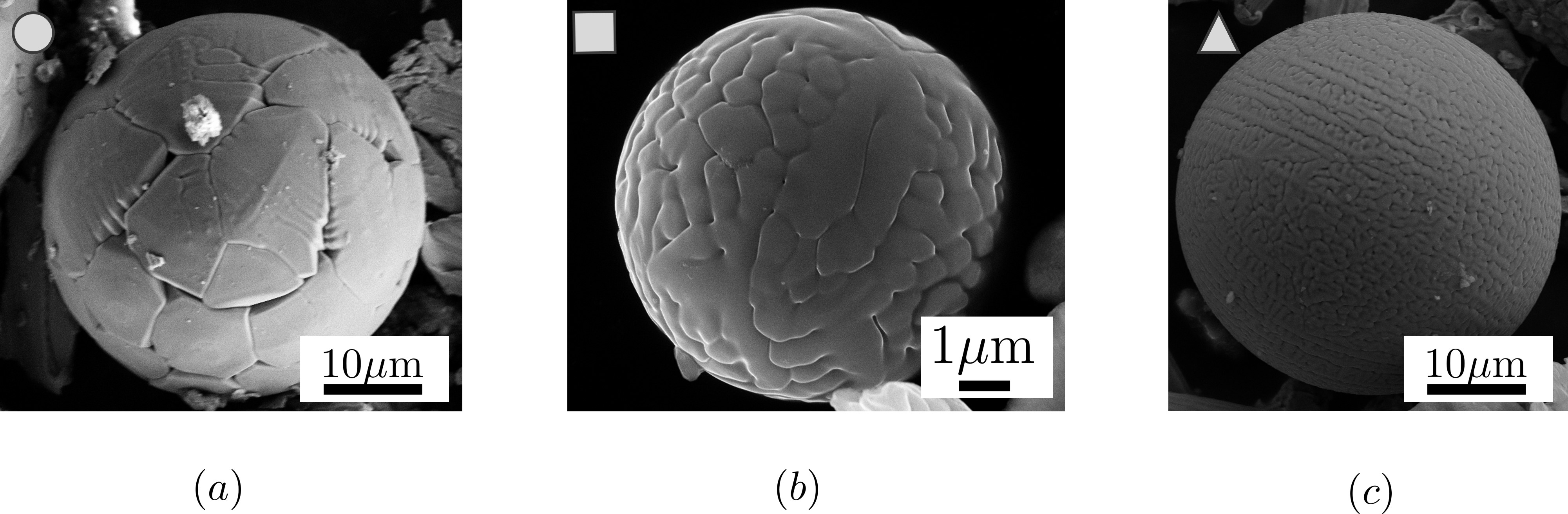}
	\caption{Morphologies observed in spherical particles of different radii. (a) Particle with radius $\approx 20\,\mu$m showing coarse dendritic structures. Panels (b) and (c) show particles of radius $\approx 4\,\mu$m and $20\,\mu$m, respectively, with different microstructural features.}
	\label{fig:morphcomparisonCandTmarker}
\end{figure}

Additional microstructures and surface morphologies to support this deduction are reproduced in Fig.~\ref{fig:morphcomparisonCandTmarker}. The panels (a) and (b) in this figure show particles of radius $\approx 20\,\mu$m and $4\,\mu$m, respectively. It can be seen that the larger particle exhibits a coarser microstructure, consistent with the corresponding deduction pertaining to \mycircle{black, fill=gray!50} and \mysquare{black, fill=gray!50} in Fig.~\ref{fig:dispandmmaxsurfacecombined}.  Likewise, comparing panels (a) and (c) in Fig.~\ref{fig:morphcomparisonCandTmarker}, we see two particles of nearly equal radius $\approx 20\,\mu$m.  These were collected from approximately the same distance from the tool-workpiece zone during abrasion, implying that they were subject to approximately similar undercooling. Interestingly, markedly different microstructures—coarse dendritic in one and fine dendritic in the other can be observed. These should then correspond to the \mycircle{black, fill=gray!50} and \mytriangle{black, fill=gray!50} markers in Fig.~\ref{fig:dispandmmaxsurfacecombined}(a) and (b) which represent two particles of similar size but varied microstructure. This implies that the initial front size, prior to instability onset, was very different in these two particles.

\begin{figure}[ht!]
  \centering
  \includegraphics[width=0.95\linewidth]{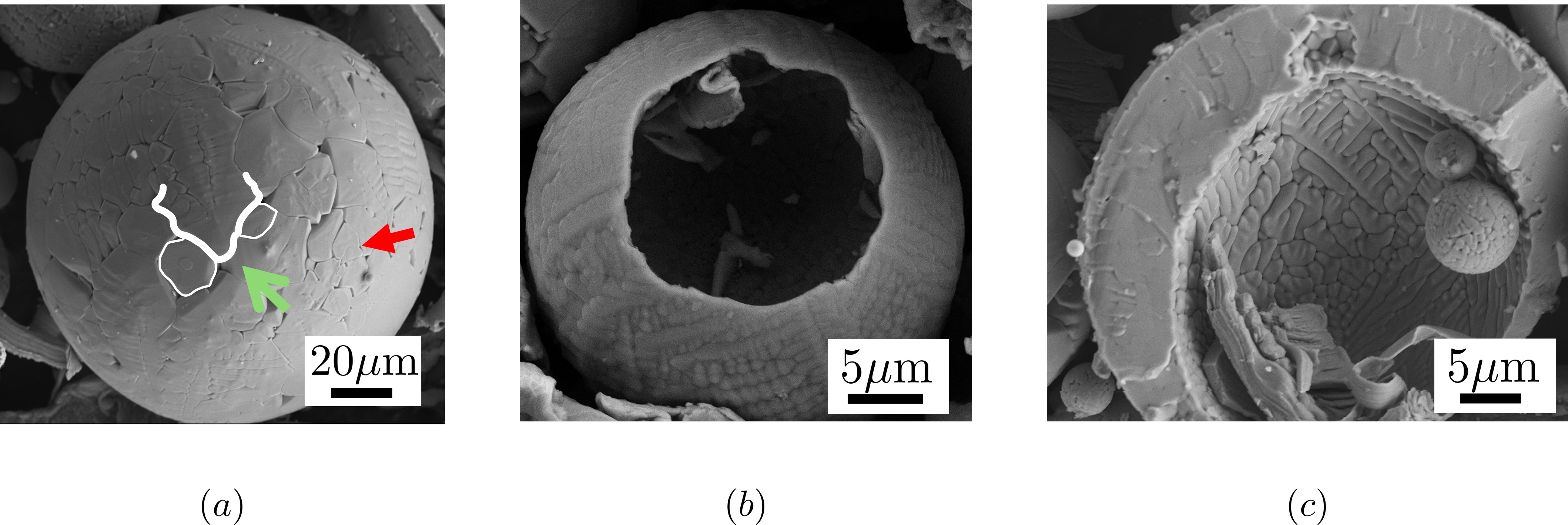}
  \caption{(a) Interaction of dendrite tip growing on the surface with a cell/dendrite tip grown radially outward. (b) A hollow spherical particle. (c) Section of a hollow particle showing dendritic structures.}
  \label{fig:corelationcombined}
\end{figure}

Finally, we present data of spherical particles that have internal voids---a tell-tale sign of the CG mode---in Fig.~\ref{fig:corelationcombined}. Panel (a) in this figure shows what appear to be dendrite arms growing on the surface and interacting with tips from the RO front (green arrow), confirming the coexistence of multiple fronts. The red arrow marks equiaxed microstructure far from surface dendrites, likely formed by the RO front reaching the outer boundary. Both Fig.~\ref{fig:corelationcombined}(b) and (c) show hollow particles and corresponding thin-shell morphologies, consistent with CG-dominated growth followed by shrinkage. This suggests that these particles were subjected to much lower undercooling, as discussed in Sec.~\ref{subsec:competingsolidificationmodes}.

In summary, the somewhat idealized growth modes used in our analysis can help explain the results of the solidification process semi-quantitatively. The interplay between nucleation mode, growth kinetics, and interface stability, modulated by droplet size and undercooling, governs the final microstructure. These results perhaps provide a framework for tailoring particle morphologies via controlled processing conditions.

\section{Discussion}
\label{sec:discussion}

Our results demonstrate that the presence of curvature fundamentally alters the mechanisms of solidification—nucleation, growth, and interface stability—in metallic droplets. In this section, we discuss the broader implications of these findings, particularly in the context of hollow particle formation and the occurrence of cellular morphologies, and highlight additional physical effects considered in our analysis.

\subsection{Alternative routes for hollow particle formation}

In addition to the CG mode, there are other potential mechanisms that could result in hollow particles, \emph{cf.} Fig.~\ref{fig:corelationcombined}(b) and (c). As discussed in Secs.~\ref{sec:experimental} and \ref{subsec:competingsolidificationmodes}, two mechanisms have been proposed in the literature for hollow particle formation during gas atomisation: (a) solidification during bag breakup of molten droplets \cite{chen2018comparative}, and (b) shrinkage and gas porosity formation~\cite{campbell1991castings, rabin1990characterization, pilch1987use}.

In the bag breakup scenario, a hollow particle forms if solidification completes before a droplet disintegrates during flight. The dimensionless breakup time $\tau$ is given by:
\begin{equation}
\tau = \dfrac{t_b V_f \bar{\rho}^{0.5}}{D},
\end{equation}
where $t_b$ is the breakup time, $V_f$ is the flow velocity, $\bar{\rho} = \rho_a / \rho_d$ is the density ratio of air to droplet, and $D$ is the droplet diameter. Using the correlation from Ref.~\cite{gel1973singularities}, $\tau = 4.5(1 + 1.2\,Oh^{1.64})$, where $Oh$ is the Ohnesorge number:
\begin{equation}
Oh = \dfrac{\mu_d}{\sqrt{\rho_d D \sigma}},
\end{equation}
with $\mu_d$ the dynamic viscosity and $\sigma$ the surface tension of the droplet. For typical material parameters, $Oh = 0.0083$ and $\tau = 4.5$, yielding $t_b \approx 0.6$\,ms. From Eq.~\ref{eqn:Rt_RO_mode}, the solidification time for a nucleus of radius $r_c^{RO} = 0.5\,\mu$m is $t_s \approx 0.3$\,ms, suggesting that bag breakup could indeed contribute to hollow particle formation. However, particles seldom remain spherical following such breakup events, in contradiction to the nearly perfect spheres commonly observed (\emph{cf.} Sec.~\ref{sec:experimental}).

Gas entrapment is another common cause of void formation. As solidification progresses, overall gas solubility decreases, leading to porosity. If the surface solidifies rapidly, forming a thin shell, the radial front may proceed inward, trapping gas. As discussed in Sec.~\ref{subsec:nucleationThermodynamics} and Fig.~\ref{fig:G_nucleation_RO_CG}, this scenario is likely for large droplets with low undercooling, where bulk nucleation is energetically unfavourable and radial growth is sluggish. Assuming radial symmetry, the volumetric shrinkage due to density differences can be estimated via mass conservation. Using $\rho_l = 7000$\,kg/m$^3$ and $\rho_s = 7800$\,kg/m$^3$ for steel (see Table~\ref{tbl:materialPropsFe}, a $30\,\mu$m droplet yields a void of $24\,\mu$m, corresponding to a shell thickness of $\approx 6\,\mu$m—consistent with Fig.~\ref{fig:corelationcombined}(c). The observed non-uniform shell thickness suggests that voids are off-centre, as also reported in optical micrographs~\cite{rabin1990characterization} and CT images~\cite{chen2018comparative}.

We finally also mention thermoelastic effects \cite{forgac1979solidification} which could result in large negative pressure inside a solidifying drop with a fully formed shell. To relieve internal stresses, the shell may buckle or deform plastically~\cite{campbell1967shrinkage, forgac1981solidification}, potentially leading to collapse and hollow particle formation.

\subsection{Occurrence of cellular morphology on spheres}

While our linear stability analysis in Sec.~\ref{subsec:stabilityGrowthFronts} cannot distinguish between dendritic and cellular morphologies, the exclusive presence of cellular structures in some particles (\emph{cf.} Fig.~\ref{fig:shapesandmorphologies}(c)) raises questions about the cellular-to-dendritic transition. Cells are defined as finger-like protrusions at the solid--liquid interface with blunt tips and no side arms, typically appearing in the intermediate regime between planar and dendritic growth ~\cite{buerger1934lineage, rutter1953prismatic}.

\begin{figure}[ht!]
  \centering
  \includegraphics[width=0.9\linewidth]{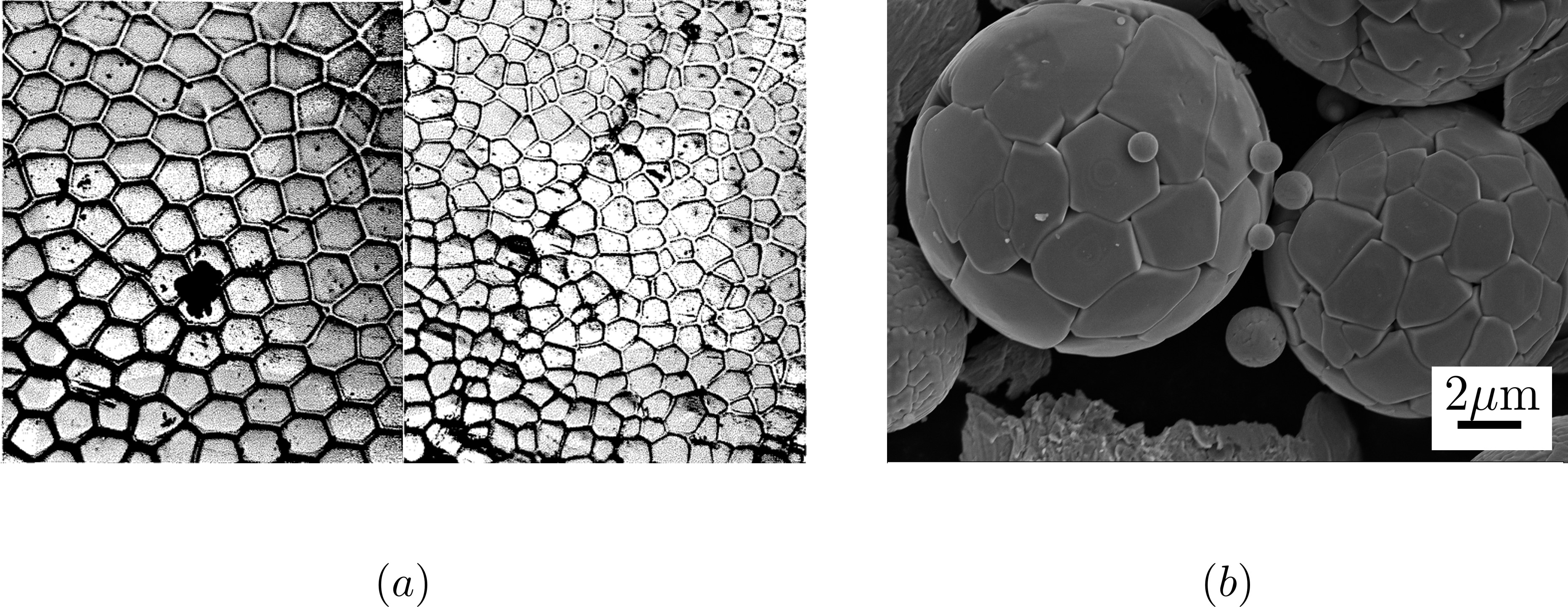}
  \caption{(a) Decanted interface of tin crystal at growth speeds of 4\,mm/min (left) and 13\,mm/min (right)~\cite{rutter1953prismatic}. (b) Spherical particles of comparable size exhibiting varying numbers of cells.}
  \label{fig:cellularcomp}
\end{figure}

Figure~\ref{fig:cellularcomp}(a) shows decanted interfaces of tin crystals at different growth rates~\cite{rutter1953prismatic}, where cell diameter decreases with increasing growth rate. Figures~\ref{fig:shapesandmorphologies}(d) and \ref{fig:cellularcomp}(b) show spherical particles with varying numbers of cells, suggesting different local growth rates. By analogy with these planar experiments, it may be speculated, based on the linear stability analysis in Sec.~\ref{subsec:stabilityGrowthFronts}, that particles such as those in Fig.~\ref{fig:cellularcomp}(b) form predominantly from an unstable RO mode. Given the small size of these spheres, the likelihood of additional instability (leading to dendrite formation) is also expected to be low.

\begin{figure}[ht!]
  \centering
  \includegraphics[width=0.9\linewidth]{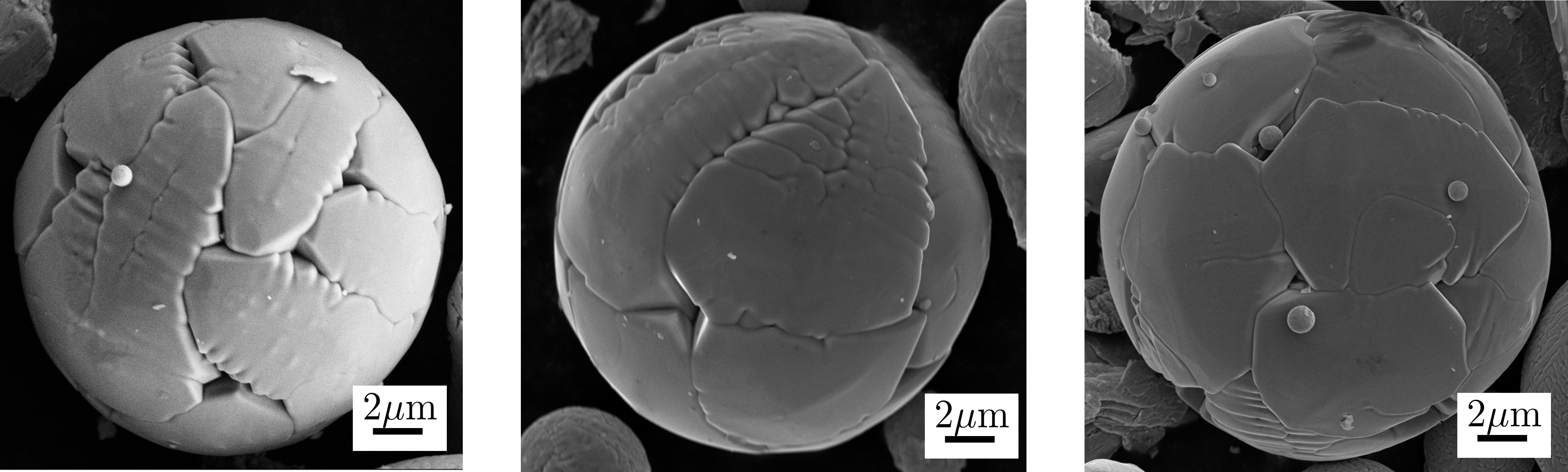}
  \caption{Spherical particles at different stages of coarsening during solidification. Remnants of dendritic side arms are visible in all cases.}
  \label{fig:coarseningofdendr}
\end{figure}

Local variations in heat transfer during the solidification process may cause different regions of the particle to experience distinct growth conditions, allowing both cells and dendrites to form and interact (see Fig.~\ref{fig:corelationcombined}(a)). As undercooled liquid becomes scarce, growth slows and the particle enters a coarsening stage. During coarsening, cells enlarge and dendrite arm spacing increases. Figure~\ref{fig:coarseningofdendr} shows particles at different coarsening stages, with remnants of dendritic side arms confirming prior dendritic growth.

\subsection{Additional effects of curvature on nucleation and growth}

As discussed in Sec.~\ref{subsec:nucleationThermodynamics}, bulk nucleation is typically considered independent of the outer boundary. However, the presence of curvature can significantly alter nucleation kinetics. While undercooling $\Delta T$ remains the primary parameter governing critical nucleus size, pressure differences across a curved boundary may modify the local temperature field around an incipient solid nucleus, with attendant effects on its thermodynamic stability.

Moreover, perfect crystal growth on a curved surface is inherently difficult due to geometric frustration. The curvature induces lattice distortions, increasing the strain energy of the growing crystal~\cite{vitelli2006crystallography, grason2016perspective}. Meng \emph{et al.}~\cite{meng2014elastic} modelled the free energy of a circular crystal growing on a sphere, incorporating a penalisation term to account for substrate curvature.

During growth, nuclei may relieve elastic frustration by forming topological defects and ramified structures~\cite{li2019ground, ortellado2022two}. In metallic systems, such defects can manifest as hot tearing or hot cracking. Campbell~\cite{campbell1991castings} defines hot tearing as a uniaxial tensile failure occurring in weak regions, often at hot spots where contraction strain accumulates. Analogously, curvature-induced strain may lead to tearing of the growing nucleus, resulting in inter-dendritic cracks. A detailed analysis of fracture patterns on solidified particles could help elucidate the local stress state during solidification.

\subsection{Consequences of the quasi-steady approximation}

The quasi-steady approximation, used in Sec.~\ref{subsec:growthKinetics}, enables closed-form expressions for growth rates and helps identify key control parameters. Physically, the Stefan number $\mathcal{S}$ represents the ratio of sensible heat to latent heat, and setting $\mathcal{S} \ll 1$ implies most thermal energy is used for phase change. This assumption tends to overestimate the interface position, providing an upper bound on solidification rate.

Numerical solutions to Stefan problems offer more accurate predictions but require continuous tracking of the moving interface, which is computationally intensive. Phase-field models, which avoid explicit interface tracking by introducing an order parameter, have been widely used to simulate solidification patterns~\cite{boettinger2002phase}. Gomez \emph{et al.}~\cite{ortellado2020phase} applied a similar approach to study curvature effects on nucleation and growth. However, their model did not couple the evolution of the order parameter with the temperature field. A fully coupled model is essential for quantitatively predicting the complex morphologies observed in curved geometries.

\subsection{Controlled solidification of spherical particles}

The abrasion-based method used in this study does not allow for controlled growth of individual particles. Ideally, the dynamics of competing growth modes should be studied independently under controlled conditions. To our knowledge, no such framework currently exists.

Freezing of water droplets on cold substrates has been investigated~\cite{chaudhary2014freezing, schremb2016solidification, jung2012mechanism}, where the substrate provides a nucleation site. Containerless solidification avoids substrate effects~\cite{herlach2012solidification}, but analyses are typically post-mortem. A framework enabling real-time observation and control of growth dynamics in spherical droplets would be invaluable. Such an approach could pave the way for defect-free particle production and deeper understanding of curvature-driven solidification phenomena.

\section{Conclusions}
\label{sec:conclusions}

This study demonstrates that the morphologies observed in spherical or near-spherical metallic particles arise from a complex interplay between growth fronts originating both within the bulk and along the surface. A generalisation of Mullins--Sekerka theory was developed to account for the influence of a finite, curved outer boundary on interface stability. This analysis revealed that the dimensionless parameter $\epsilon \propto 1/a$—representing the ratio of interfacial perturbation wavelength to particle size—emerges as a key factor governing morphological evolution.

The competition between radial outward (RO) and circumferential growth (CG) modes is dictated by the ratio of initial seed size to particle size, with the dominant mode ultimately determining the final microstructure. Under prevailing solidification conditions, both dendritic and cellular morphologies can grow independently. Variations in initial nucleation conditions across particles lead to a distribution of growth behaviours and interaction patterns. Furthermore, post-solidification coarsening processes allow dendrites and cells to coexist within the same particle, resulting in the diverse morphologies observed experimentally.

These findings provide a unified framework for interpreting microstructural features in rapidly solidified spherical particles and offer insights into how curvature, undercooling, and growth kinetics collectively shape the final morphology.

\bibliography{bibfile}
\bibliographystyle{vancouver}

\end{document}